\documentclass[11pt,a4paper]{article}

\usepackage{jheppub}
\usepackage{amsmath}
\usepackage{hyperref}
\usepackage{amssymb}
\usepackage{mathtools}
\usepackage{mathrsfs}
\usepackage{amsfonts}
\usepackage{bbm}
\usepackage{enumerate}
\usepackage{array}
\usepackage{booktabs}
\usepackage{textcomp}
\usepackage{caption}
\usepackage{cite}
\usepackage{cancel}
\usepackage{units}
\usepackage{multirow}
\usepackage{pst-all}
\usepackage{longtable}
\usepackage{relsize}
\usepackage[]{natbib}

\title{Holography and Conformal Anomaly Matching}

\author[a]{Alejandro Cabo-Bizet}
\author[b]{Edi Gava}
\author[c]{K.S. Narain}

\affiliation[a]{SISSA and INFN,
Via Bonomea 265, 34128 Trieste, Italy}
\affiliation[b]{INFN, sezione di Trieste, Italy}
\affiliation[c]{ICTP,  Strada Costiera 11, 34014 Trieste, Italy}

\emailAdd{acabo@sissa.it}
\emailAdd{gava@ictp.it}
\emailAdd{narain@ictp.it}

\begin{document}

\abstract{ We discuss various issues related to the understanding of the conformal anomaly matching in CFT
from the dual holographic viewpoint.  First, we act with a  PBH diffeomorphism on a generic 5D RG flow geometry
and show that the corresponding on-shell bulk action reproduces the Wess-Zumino term for the dilaton of broken conformal symmetry, with the expected coefficient $a_{UV}-a_{IR}$.  Then, we consider a specific 3D example of  RG flow whose UV
asymptotics is normalizable and admits a 6D lifting. We promote a modulus $\rho$ appearing
in the geometry to a function of boundary  coordinates. In a 6D description $\rho$ is the scale of an $SU(2)$ instanton.
We  determine the smooth deformed background up to second
order in the space-time derivatives of $\rho$ and  find that the 3D on-shell action reproduces a boundary kinetic term for the
massless field $\tau=\log \rho$ with the correct coefficient $\delta c=c_{UV}-c_{IR}$. We further analyze the linearized
fluctuations around the deformed background geometry  and  compute the one-point functions $<T_{\mu\nu}>$
and show that they are reproduced by a Liouville-type  action for the massless scalar $\tau$, with background charge due to the coupling to the 2D curvature $R^{(2)}$. The resulting central charge
matches $\delta c$. 
We give an interpretation of this action in terms of the $(4,0)$ SCFT of the D1-D5 system in type I theory.}

\arxivnumber{}

\maketitle

\flushbottom

\section{Introduction}
\label{sec:Introduction}

The proof of the a-theorem in D=4 CFT and the alternative  proof of c-theorem in D=2 CFT \citep{Zamolodchikov}, given in \citep{Komargodski1, Komargodski2}, inspired by the anomaly matching argument of \citep{Schwimmer1}, has prompted several groups
to address the issue of a description of the corresponding mechanism
on the dual gravity side \citep{TheisenMyers, Hoyos}. While a sort of $a(c)$-"theorem " is known to hold  for RG-flows in the context of gauged supergravity  \citep{Freedman1,Freedman2},
as a consequence of the positive energy condition, which
guarantees the monotonic decrease  of the $a(c)$ function from UV to IR \citep{Sinha1}\footnote{ Different approaches  have been discussed lately \citep{Casini,Ryu,Rattazzi}.}, one of the aims of the renewed interest
on the topic has been somewhat different: the field-theoretic anomaly matching argument implies the existence of an IR effective action for the conformal mode, which
in the case of spontaneous breaking of conformal invariance is the physical dilaton, whereas for a RG flow due to relevant perturbations is
a Weyl mode of the classical background metric ("spurion"). In any case, upon combined Weyl shifting of the conformal mode and the background metric, the effective action
reproduces the conformal anomaly of amount $a_{UV}-a_{IR}$ ($c_{UV}-c_{IR}$), therefore matching the full conformal anomaly of the UV CFT.
This effective action therefore is nothing but the Wess-Zumino local term corresponding to broken conformal invariance. So, one obvious
question is  how to obtain the correct Wess-Zumino term for the dilaton (or spurion) from the dual gravity side. One of the purposes of the
present paper is to discuss this issue offering a different approach from those mentioned above.
In known examples of 4D  RG flows corresponding to spontaneous breaking of conformal invariance on the Coulomb branch of $N=4$ Yang-Mills
theory \citep{Bianchi1,Bianchi2,Freedman3,Girardello1}, indeed the existence of a massless scalar identifiable with the CFT's dilaton (see also \citep{Hoyos, Bajc}) has been shown.  However,
the background geometry is singular in the IR, so that one does not have a full control on the geometry all along the RG flow. It would be therefore
desirable to have an explicit example which is completely smooth from UV to IR, and indeed we will discuss such an example in the $AdS_3/CFT_2$ context.

Before going to analyze in detail a specific example, we will generally ask what is the bulk mode representing the spurion field of the CFT. The spurion couples to field theory operators in according to their scale dimension and  transforms under conformal transformations  by Weyl shifts. These properties point towards an identification of this mode with the PBH (Penrose-Brown-Henneaux)-diffeomorphism, which are bulk diffeomorphisms inducing
Weyl transformations on the boundary metric, parametrized by the spurion field $\tau$. This identification has been first
adopted  in \citep{Imbimbo,Schwimmer2} to study holographic conformal anomalies and also recently in \citep{TheisenMyers,Hoyos,Sinha2} to address the anomaly matching
issue from the gravity side.


As will be shown in section \S  2, for the  case of a generic 4D RG flow,  by looking at how PBH diffeomorphisms act
on the background geometry at the required order in a derivative expansion  of $\tau$, we will compute the regularized bulk action for the PBH-transformed
geometry and show that it contains a finite contribution proportional to the Wess-Zumino term for $\tau$,  with proportionality constant given by $a_{UV}-a_{IR}$.


In the case where conformal invariance is spontaneously broken, when $D>2$, one expects to have a physical massless scalar on the boundary CFT,
the dilaton, which is the Goldstone boson associated to the broken conformal invariance. As stressed in \citep{Hoyos}, one expects on general grounds
that the dilaton should be associated to a normalizable bulk zero mode, and therefore cannot be identified with the PBH spurion, which is
related to a non normalizable deformation of the background geometry.

In section \S 3 we will follow a different approach to the problem: starting from an explicit, smooth RG flow geometry in 3D gauged supergravity \citep{Parynha},
we will promote some  moduli appearing in the solution to space-time dependent fields. More specifically, we will identify a modulus which, upon lifting the solution to 6D,
is in  fact the scale $\rho$ of an $SU(2)$  Yang-Mills instanton. We will then find the new solution of the supergravity equations of motion up to second order
in the space-time derivatives  of $\rho$. We will find that demanding regularity of
the deformed geometry forces to switch on a source
for a scalar field.  We will then compute the on-shell bulk action
and verify that this reproduces the correct kinetic term boundary action for the massless scalar field $\tau=\log\rho$, with coefficient $\delta c=c_{UV}-c_{IR}$ \footnote{This new field $\tau$ should not be confused with the spurion fields discussed in sections \S 2 and \S 3. We hope not to confuse with this abuse of notation.}.
The computation of the CFT effective action is done up to second order in derivative expansion. Namely, only the leading term in the full IR effective action is computed, and our procedure is similar to the one followed in \citep{Shiraz}  for the derivation of the equations of hydrodynamics from
AdS/CFT.
In section \S 4 we reconsider the problem from a 6D viewpoint \citep{Nishino}: the 6D description has the advantage of
making more transparent the $10D$ origin of our geometry in terms of  a configuration of $D1$ and $D5$ branes in type I string theory
\footnote{As it will be clear in section \S 3, the background geometry involves to a superposition of $D5$ branes and a gauge 5-brane \citep{Callan} supported
by the $SU(2)$ instanton. The latter is interpreted as a $D5$ branes in the small instanton limit $\rho\rightarrow 0$\citep{WittenSI}.}.


Here we  take one step further: not only we determine the deformed background involving
two derivatives of $\rho$ but also  solve the linearized equations of motion  around it to determine the on-shell fluctuations. This  allows
us to compute one-point functions of the boundary stress-energy tensor $<T_{\mu\nu}>$, from which we  deduce that the boundary
action for $\tau$ is precisely the 2D Wess-Zumino term of broken conformal invariance, i.e. a massless scalar coupled to
the 2D curvature $R^{(2)}$ and overall coefficient  $\delta c$.  An obvious  question is  what the field  $\tau$ and its action  represent on the dual CFT.
We will argue that the interpretation of the effective field theory for $\tau$ is a manifestation of the
mechanism studied in \citep{SW}, describing  the separation of a $D1/D5$ sub-system from a given  $D1/D5$ system from the viewpoint  of the
 $(4,4)$ boundary CFT. There, from the Higgs branch, one obtains an action for the radial component of vector multiplet scalars which
 couple to the hypermultiplets,  in the form of a 2D scalar field
 with background charge, such that its conformal anomaly compensates the variation of the central charge due
 to the emission of the sub-system.
 In our case we will see that in the limit $\rho\rightarrow \infty$, the gauge five-brane
 decouples,  whereas in the limit $\rho\rightarrow 0$ it becomes a D5-brane: these two limits correspond in turn to the IR and UV regions of
 the RG flow, respectively.  The effective action for $\tau=\log\rho$ accounts, in the limit of large charges, precisely for the $\delta c$
 from the UV to the IR in the RG flow. We will give an interpretation of  the action for $\tau$ in terms of the effective field
 theory of the D1-D5 system in presence of D9 branes in type I theory.

We stress that the above procedure, although, for technical reasons,  implemented explicitly in the context of an $AdS_3/CFT_2$ example,
we believe should produce the correct Wess-Zumino dilaton effective action even in the $D=4$ case, had we an explicit, analytic and smooth RG flow triggered by a v.e.v. in the UV. Of course, in this case we should have pushed the study of equations of motion up to fourth order in the derivative expansion.

\newpage
\section{The Holographic Spurion}
The aim of this section is to verify that the quantum effective action for the holographic spurion in 4D contains the Wess-Zumino term,
a local term whose variation under Weyl shifts of the spurion field reproduces the  conformal anomaly \footnote{This is a combination of a Weyl shift of the background metric with a compensating shift in the spurion field. In this way the remaining variation is independent of the spurion field. It depends only on the background metric.}, with coefficient given by the difference
of UV and IR $a$-central charges, in accordance with the anomaly matching argument.
We start by characterizing a generic RG flow  background and the action of PBH diffeomorphisms on it.
The action of a special class of PBH diffeomorphisms introduces a dependence of the background
on a boundary conformal mode which will play the role of the spurion.
Indeed,  we will  verify that  the corresponding on-shell Einstein-Hilbert action gives the correct Wess-Zumino term for the conformal mode introduced through PBH diffeo's.
We then study the case of a flow induced by a dimension $\Delta=2$ CFT operator, and check that boundary contributions coming from the Gibbons-Hawking term and counter-terms do not affect the bulk result. A derivation of the Wess-Zumino action  has appeared in \citep{Sinha2},
studying pure gravity in AdS in various dimensions: the spurion $\phi$  is introduced as deformation of the UV cut-off boundary surface from $z$ constant to $z=e^{\phi(x)}$,
$z$ being the radial coordinate of AdS. In appendix \ref{AnomalyM} we present a covariant approach to get the same result for the WZ term.

\subsection{Holographic RG flows}
We start by characterizing a generic RG flow geometry. For the  sake of simplicity, we are going to work only with
a single scalar minimally coupled to gravity. In the next section we will consider a specific example
involving two scalar fields.
The action comprises the Einstein-Hilbert term, the kinetic and potential terms for a scalar field $\phi$,
and the Gibbons-Hawking extrinsic curvature term at the boundary of the space-time manifold $M$:
\begin{align}\label{action}
S=\int_{M} d^{d+1}x\sqrt{G}(\frac{1}{4}R+(\partial \phi)^2-V(\phi))-\int_{\partial M}d^dx\sqrt{\gamma}\frac{1}{4}2K ,
\end{align}
where $K$ is the trace of the second fundamental form,
\begin{align}
2K=\gamma^{\alpha \beta}L_{n}\gamma_{\alpha \beta},
\end{align}
and $\gamma$ is the induced metric on the boundary of $M$, $\partial M$, $L_{n}$ is the Lie derivative with respect to the unit vector
field $n$ normal to $\partial M$.

 The metric has the form:
\begin{align}\label{metric1}
ds^2=\frac{l^2(y)}{4}\frac{dy^2}{y^2}+\frac{1}{y}g_{\mu \nu}(y)dx^\mu dx^\nu,
\end{align}
which is an $AdS_5$ metric for constant $l(y)$ and $g_{\mu \nu}(y)$ ($\mu, \nu=0,1,2,3.$). A RG flow geometry is then
characterized by the fact that the above geometry is asymptotic to $AdS_5$
both in the UV and IR limits,  $y\rightarrow 0$ and $y\rightarrow\infty$, respectively.

We assume that  the potential $V(\phi)$ has two $AdS_5$ critical points that we call $\phi_{UV(IR)}$ and the background
involves a  solitonic field configuration $\phi(y)$ interpolating monotonically between these two critical points:
\begin{align}
\phi(y)&\sim \delta \phi(y)+\phi_{UV}, \ \text{when} \ y\rightarrow \infty \\
\phi(y)&\sim \delta \phi(y)+\phi_{IR}, \ \text{when} \ y\rightarrow 0
\end{align}
Around each critical point there is an expansion:
\begin{align}\label{Potential}
V(\phi)\sim\Lambda_{UV(IR)}+m_{UV(IR)}^2{\delta \phi(y)}^2+ o({\delta \phi(y)}^4),
\end{align}
where $\delta \phi(y)= \phi(y)-\phi_{UV(IR)}$. By using (\ref{Potential}) in the asymptotic expansion of the equations of motion:
\begin{align}\label{eqmotions}
\frac{1}{4}R_{\mu \nu }=\partial _{\mu }\phi \partial _{\nu }\phi +\frac{1}{3
}V[\phi ],
\end{align}
one sees that the constants $\Lambda_{UV(IR)}$ play the role of cosmological constants and fix also the radii of the two $AdS_5$'s.

We discuss here the possibility to work in a gauge that makes easier to appreciate how only the boundary data is determining
the spurion effective action.
Consider a RG flow geometry of the form (\ref{metric1}). Poincar\'{e} invariance of the asymptotic value of the metric implies $g_{\mu\nu}(y)=g(y)\eta_{\mu \nu}$. This is going to be an important
constraint later on. The scale length function $l^2(y)$ has the following  asymptotic behaviour:
\begin{align}
l^2(y_{UV})& \sim L^{2}_{UV}+\delta l_{UV}y_{UV}^{n_{UV}}, \ L^2(y_{IR})\sim L^{2}_{IR}+\frac{\delta l_{IR}}{y_{UV}^{n_{IR}}}.
\end{align}
Notice that there is still the gauge freedom:\[(x,y) \rightarrow (x,h \times y),\] where $h=h(y)$ is any smooth function with asymptotic values 1 in
the UV/IR fixed points. This gauge freedom allows to choose positive integers $n_{UV}$ and $n_{IR}$ as large as desired. In particular it is always possible
to choose $n_{UV}>2$. This gauge choice does not change the final result for the effective action because this is a family of proper
diffeomorphisms leaving invariant the Einstein-Hilbert action (we will comment on this fact later on). Its use is convenient in order to make clear how only leading behaviour in the background solution is relevant  to our computation. At the same time  it allows to get rid of any back-reaction of $\delta l_{UV}$ and $\delta l_{IR}$ in the leading  UV/IR asymptotic expansion of the equations of motion.
The metric $g_{\mu\nu}$ has the following UV expansion, for $y\rightarrow 0$:
\begin{align}\label{NToBExpansion1}
g_{\mu\nu}=g_{\mu\nu}^{(0)}+g_{\mu\nu}^{(2)}y+y^2\left(g_{\mu\nu}^{(4)}+h_{\mu\nu}^{(4)}\log(y)+{\tilde h}_{\mu\nu}^{(4)} \log^2(y)\right)+o(y^{3}) ,
\end{align}
and a bulk scalar field dual to a UV field theory operator of conformal dimension $\Delta=2$ that we denote as $\textbf{O}_{(2)}$, behaves like:
\begin{align}\label{}
\delta \phi=\phi^{(0)}y+\widetilde{\phi}^{(0)}y \log(y)+...,
\end{align}
where the $...$ stand for UV subleading terms. From the near to boundary expansion of the Klein-Gordon equations one reads the useful relation
between the conformal weight of $\textbf{O}_{(2)}$ and the mass of $\phi$ on dimensional $AdS_{d+1}$:
\begin{align}
\Delta_{UV}=\frac{d}{2}+\sqrt{\frac{d^2}{4}+m^2 L_{UV}^2}.
\end{align}
In this critical case we have  the standard relation between asymptotic values of bulk fields and
v.e.v.'s or  sources for the dual CFT operators:  namely $\phi_{(0)}$ is the v.e.v. and $\tilde{\phi}_{(0)}$ the source.
We have chosen the case $\Delta=2$ to take a particular example, but one can easily generalize the results  to any other value of $\Delta\leq4$. In the remaining of the section we refer only to relevant perturbations.

\subsection{On the PBH diffeomorphisms}


 The PBH diffeomorphisms transform, by definition,
 the line element (\ref{metric1}) into:
\begin{align}\label{lineelement2}
ds^2 =\frac{l^2(e^{\tau} y) }{4 y^2}dy^2+\frac{1}{y}\tilde{g}_{\mu\nu}(y)dx^{\mu}dx^{\nu} ,
\end{align}
with $\tilde{g}_{\mu\nu}$ given by an UV asymptotic expansion of the form (\ref{NToBExpansion1}):
\begin{align}\label{NToBExpansion2}
\tilde{g}_{\mu\nu}=e^{-\tau}g_{\mu\nu}^{(0)}+... \,
\end{align}
and $h^{(4)}_{1,2}$ and $g^{(i)}$, with $i=2,4$, determined in terms of the boundary data by the near to boundary expansion of the equations of
motion (\ref{NearToBEoM}).

For the static RG flow geometry at hand, (\ref{metric1}) a PBH transformation has the following structure in terms of derivatives of $\tau$:
\begin{eqnarray}
\begin{split}\label{PBH}
x^{\mu}\rightarrow {x^{\tau}}^{\mu}= x^{\mu}-a^{(1)}[e^{\tau}y]\partial^{\mu}\tau-&{}a^{(2)}[e^{\tau}y]\partial_{\mu}\Box \tau -a^{(3)}[e^{\tau}y]\Box^{\mu \nu}
\tau\partial_{\nu}\tau - a^{(4)}[e^{\tau}y]\Box \tau
\partial^{\mu}\tau \\& \qquad\qquad \qquad \qquad\qquad \qquad \qquad- a^{(5)}[e^{\tau}y](\partial \tau)^2 \partial^{\mu}\tau+O\left(\partial^5\right),  \\
y \rightarrow y^{\tau}= y e^{\tau}+b^{(1)}[e^{\tau}y](\Box\tau)+&{}b^{(2)}[e^{\tau}y](\Box\Box\tau)+ b^{(3)}[e^{\tau}y](\partial\tau)^2 +
b^{(4)}[e^{\tau}y](\Box\tau)(\partial\tau)^2\\
&\qquad +b^{(5)}[e^{\tau}y](\partial\tau)^4+b^{(6)}[e^{\tau}y]\partial_{\mu}\tau \Box^{\mu \nu}\tau\partial_\nu\tau+b^{(7)}[e^{\tau}y](\Box \tau)^2\\&\qquad \qquad \qquad \qquad \qquad \qquad \qquad+b^{(8)}[e^{\tau}y]\partial_{\mu}(\Box \tau)\partial^{\mu}\tau+O(\partial^{6}),\end{split}
\end{eqnarray}
where the $...$ stand for higher derivative in $\tau$ dependence. Covariant indices are raised up with the metric
$g^{\mu\nu}(y)=g^{-1}(y)\eta^{\mu \nu}$. Notice we have written the most general boundary covariant form and that this IR expansion of the full transformation is valid along the full flow geometry up to the IR cut off, 
not only  near to boundary. The constraints implied by preserving the form  (\ref{lineelement2})
allow to determine the form factors $a^{(i)}$ and $b^{(i)}$ in terms of the scale length function $l$.
To begin with, it is immediate to see that :\[\partial_z a^{(1)}=\frac{l^2(z)}{4},\] where $z=e^{\tau}
y$, which can be readily integrated.
Some of these form factors can be settled to zero without lost of generality, since  they are solution of homogeneous differential equations.
Let us study the following one $b^{(1)}$. We can look at second order in derivatives contribution of $\delta x^{\mu}\equiv {x^{\tau}}^{\mu}- x^{\mu} $ to the ($y,y$)
component of the metric, which is $\sim (\partial \tau)^2$. The contributions coming from $\delta y\equiv y^{\tau}-y $ contains a linear order in $y$ term proportional to
\[ \left(\left(-\frac{l}{z}+\partial_{z}l\right) b^{(1)}+l \partial_y b^{(1)}\right)\Box \tau,\]
that does not match any contribution from $\delta x^{\mu}$ and also a term proportional to $(\partial \tau)^2$. This implies $b^{(1)}$ has to be
taken to vanish. Consequently $a^{(2)}$ would vanish.
In the same fashion one can prove $b^{(2)}$ can be taken to vanish and $b^{(3)}$ can be found to obey the following inhomogeneous first order
differential equation:\[\left(\frac{\partial_z l}{l}- \frac{1}{z}\right)b^{(3)}+\partial_z b^{(3)}=-\frac{l^2}{8} z ,\] which can
be solved asymptotically to give: \begin{align}
b^{(3)}&\sim -\frac{L_{UV}^2}{8}z^2+O\left(z^{n_{UV}+2}\right), \ b^{(3)}\sim-\frac{L_{IR}^2}{8}z^2+O\left(z^{-n_{IR}+2}\right).
\end{align}
Notice that so far, we have always taken the trivial homogeneous solution. In fact we are going to see that this choice corresponds to the minimal description of the spurion. The choice of different PBH representative \footnote{ Namely, to pick up non trivial solutions of the homogeneous differential equations for the form factors. } would translate in a local redefinition of the field theory spurion. In the same line of logic one can find that:
\begin{align}
\partial_z a^{(5)}=\frac{l^2}{4} \partial_z b^{(3)}, \ \partial_z a^{(3)}=\frac{l^2}{2} \frac{b^{(3)}}{z}, \ \partial_z a^{(4)}=0.
\end{align}
From these we can infer that $b^{(4)}$, $b^{(7)}$ and $b^{(8)}$ obey homogeneous differential equations provided $a^{(4)}$ is taken to vanish, so we set them to zero too. The
following constraints:
\begin{align}
\left( \left(\frac{\partial_{z}l}{l}-\frac{1}{z}\right) b^{(5)}+\partial_z b^{(5)}\right)&=-\left(\frac{(\partial_z l)^2+l\partial^2_z l}{2
l^2}+\frac{3}{2}\frac{1}{z^2} \right)\left(b^{(3)}\right)^2\\&\qquad-\left(2\left(\frac{\partial_z l}{l}-
\frac{1}{z}\right)b^{(3)}+\frac{1}{2}\partial_z b^{(3)}\right)\partial_z b^{(3)}-\frac{l^2}{4}z \partial_z b^{(3)},\\
\left( \left(\frac{\partial_{z}l}{l}-\frac{1}{z}\right) b^{(6)}+\partial_z b^{(6)}\right)&=-\frac{l^2}{2} b^{(3)},
\end{align}
give the UV/IR asymptotic expansions for the form factors:
\begin{align}
b^{(5)}&\sim-\frac{L_{UV}^4}{128} z^3+...,\ b^{(5)}\sim-\frac{L_{IR}^4}{128}
z^3+...,\\
b^{(6)}&\sim-\frac{L_{UV}^4}{32} z^3+...,\  b^{(6)}\sim-\frac{L_{IR}^4}{32} z^3+...,
\end{align}
where the $...$ stand for subleading contributions. In appendix (\ref{AppendixA}) we extend these results to the case of non static geometries. We use
those non static cases in section \S 3 to check out the general results of this section in a particular example.

Before closing the discussion let us  comment about a different kind of PBH modes. To make the discussion simpler we restrict our analysis to the level of PBH zero modes i.e.  $\tau$ is taken to be a constant. Then, is easy to see that one can take the transformation \[y\rightarrow y^{\tau}= e^{h \times \tau}  y, \ \text{with} \ h(y)\xrightarrow[y\rightarrow (0, \infty)]{} h_{(UV,IR)}. \] This arbitrary function $h$ constitutes a huge freedom. In particular we notice that one can choose a PBH which does not affect the UV boundary data at all, but does change the IR side, namely such that: \[h\sim 0,\ h\sim 1 \] respectively, or vice versa. This kind of  PBH's are briefly considered in appendix \ref{AppendixA}.  



Besides acting on the metric the change of coordinates also changes the form of the scalars in our background. We focus on the UV asymptotic. So, for instance the case of the dual to a $\Delta=2$ operator:
\begin{align}\label{vev}
 \widetilde{\phi}^{(0)}\rightarrow e^{\tau}\widetilde{\phi}^{(0)},\ {\phi}^{(0)}\rightarrow e^{\tau}{\phi}^{(0)}+\tau
 e^{\tau}\widetilde{\phi}^{(0)}.
\end{align}
Notice the source transforms covariantly,  but not the v.e.v.. This asymptotic action will be useful later on when solving the near to boundary equations of motion. 

As already mentioned, we assume smoothness of the scalar field configurations in the IR. It is interesting however to explore an extra source of IR divergencies. The original 5D metric is assumed to be smooth and asymptotically AdS in the IR limit, $y \rightarrow \infty $:
\begin{align}\label{metricIR}
ds_{IR}^{2}=\frac{L_{IR}^{2}}{4}\frac{dy ^{2}}{y ^{2}}+\frac{1}{y }g_{\mu
\nu }^{(0)}dx^{\mu }dx^{\nu }.
\end{align}
This AdS limit assumption implies that $g^{(0)}_{\mu \nu}=\eta_{\mu \nu}$. Non trivial space time dependence for $g_{(0)}$ sources an infinite tower of extra contributions that break AdS limit in the IR. For instance, a Weyl shifted representative will alter the IR AdS behaviour. The change is given by:
\[
g_{\mu \nu }^{(0)}\rightarrow e^{-\tau }g_{\mu \nu }^{(0)}+y
g_{\mu\nu}^{(2)}[e^{-\tau }g^{(0)}]+y ^{2}g_{\mu\nu}^{(4)}[e^{-\tau }g^{(0)}]+y ^{2}(h_{\mu\nu}^{(4)}[e^{-\tau }g^{(0)}]\log (y ))+...,
\]
in (\ref{metricIR}). Clearly AdS IR behaviour, $y\rightarrow \infty$, is broken in this case. This is related with the fact that PBH diffeomorphisms are singular changes of coordinates in the IR. These modes alter significantly the IR behaviour of the background metric.



Let us comment  on a different approach that will be employed in the following to
study the effect of PBH  diffeo's. Clearly PBH  diffeo's map a solution of the EoM 
into another solution. 
By knowing the UV and IR leading behaviours, one could then use near to boundary equations of motion to reconstruct next to leading behaviour in both extrema of the flow. Namely we can find the factors $g^{(2)}$, $g^{(4)}$ and $h^{(4)}$'s in (\ref{NToBExpansion1}) in terms of the Weyl shift of the boundary metric $e^{\tau}g^{(0)}$. We can then evaluate the bulk and boundary GH terms of the action with this near to boundary series expansion. Some information will be unaccessible with this approach, concretely the finite part of the bulk term remains unknown after use of this method. In appendix \ref{NTBExpansion} we compute the divergent terms of the bulk term and find exact agreement
with the results posted in the next subsection.  We will use this procedure to evaluate the GH and counter-terms indeed.

\subsection{Wess-Zumino Term}
Given its indefinite $y$-integral  $S[y]$, the bulk action can be written as:
\[S_{bulk}=S[y_{UV}]-S[y_{IR}].\]
The divergent parts of the bulk action come from the asymptotic expansions of the primitive $S$:
\begin{align}\label{NTBAction}
S \sim \int d^4x\left( \frac{a_{UV}^{(0)}}{y _{UV}^{2}}+\frac{a_{UV}^{(2)}}{y
_{UV}}+a_{UV}^{(4)}\log (y_{UV})+O(1)\right), \ S \sim S_{IR}.
\end{align}
For a generic static RG flow solution $a^{(0)}_{uv,ir}=\frac{1}{2 L_{UV/IR}}$. The factors $a^{(2)}_{UV/IR}$ and $a^{(4)}_{UV/IR}$, will depend on the specific matter content of the bulk theory at hand. As for our particular choice of $\Delta$'s in the UV/IR, the $a^{(2)}_{UV/IR}$ coefficients are proportional to the 2D Ricci Scalar $R$ of the boundary metric $g^{(0)}$ and vanish for the static case $g^{(0)}_{\mu \nu}= \eta_{\mu \nu}$ (See equation (\ref{volumeFactor}), and (\ref{as}).). However, a different choice of matter content could provide a non trivial $a^{(2)}_{UV/IR}[\eta_{\mu \nu}]$ dependence on the parameters of the flow, so in order to keep the discussion as general as possible until the very end of the section we keep the static limit of both $a^{(2)}_{UV/IR}$  as arbitrary. As for the expansions of the primitive $S$ in a generic static case, one gets thence:
\begin{align}
S[y_{UV}]\sim \int d^4 x \left( \frac{1}{2 L_{UV}} \frac{1}{y_{UV}^2}+\frac{a_{UV}^{(2)}[\eta_{\mu \nu}]}{y
_{UV}}+a_{UV}^{(4)}[\eta_{\mu \nu}]\log (y_{UV})+O\left(1\right)\right), \\
S[y_{IR}] \sim \int d^4 x \left( \frac{1}{2 L_{IR}} \frac{1}{y_{UV}^2}+\frac{a_{IR}^{(2)}[\eta_{\mu \nu}]}{y
_{IR}}+a_{IR}^{(4)}[\eta_{\mu \nu}]\log (y_{IR})+O\left(1\right)\right).
\end{align}
The terms $a_{uv, ir}^{(4)}[\eta_{\mu \nu}]$ are the contributions to the Weyl anomaly coming from the matter sector of the dual CFT, they must be proportional to the sources of the dual operators. The order one contribution is completely arbitrary in near to boundary analysis. Notice that we have freedom to add up an arbitrary, independent of y functional, $\int d^4 x \  C$, in the expansions. The difference of both of these functionals carries all the physical meaning and it is undetermined by the near to boundary analysis. To determine its dependence on the parameters of the flow,  full knowledge of the primitive $S$ is needed.

Next we aim to compute the change of the bulk action introduced before,
under an active PBH diffeomorphism. The full action is invariant under (passive)  diffeomorphisms
$x^\mu=f^\mu(x')$, under which, for example,  the metric tensor changes as:
\begin{align}\label{GCT}
 g'_{\mu\nu}(x')=\left(\frac{\partial x^{\rho} }{\partial x'^{\mu}}\right)\left(\frac{\partial x^{\sigma}}{\partial x'^{\nu}}\right)g_{\rho \sigma}(x),
\end{align}
and similarly for other tensor fields. Here the transformed tensors are evaluated at the new coordinate $x'$.
On the other hand by an active diffeomorphism, the argument of a tensor field is kept fixed, i.e:
\begin{align}
g(x)\rightarrow g'(x).
\end{align}
The infinitesimal version of this transformation above is given by the Lie-derivative acting on $g$.
%
The difference between the two viewpoints becomes apparent on a manifold $M$ with boundaries. Let us take a manifold with two disconnected boundaries to be time-like hypersurfaces.  An integration of a scalar density  over this manifold is invariant in the following sense:
\begin{align}
S[B_{UV},B_{IR},g]&=\int^{B_{UV}}_{B_{IR}}d^Dx\sqrt{g(x)} L[g(x)]\\ &= \int^{f^{-1}(B_{UV})}_{f^{-1}(B_{IR})}d^Dx'\sqrt{g'(x')} L[g'(x')]\\&=S[f^{-1}(B_{UV}),f^{-1}(B_{IR}),g'],\label{JustifiPBH}
\end{align}
where the boundaries are denoted by $B_{UV(IR)}$. By $f^{-1}({B_{UV}})$ we mean the shape of the boundaries in the new coordinates $x^\prime=f(x)$.
On the other hand, under an active transformation we have the change:
\begin{align}
S[B_{UV},B_{IR},g]\rightarrow S[B_{UV},B_{IR},g'].
\end{align}
By using (\ref{JustifiPBH}), the variation of the corresponding functional under an active diffeomorphism can be written as:
\begin{align}
\Delta_fS&=S[B_{UV},B_{IR},g']-S[{B_{UV}},{B_{IR}},g]\nonumber\\
&=S[f(B_{UV}),f(B_{IR}),g]-S[{B}_{UV},{B}_{IR},g],\label{VarAction}
\end{align}
where in the last step we have used the invariance under the passive diffeomorphism induced by the inverse map $f^{-1}$. Of course, if the maps $f$ or $f^{-1}$ leave invariant the boundary conditions then the functional $S$ is invariant even under the active transformation induced by them.

From now on in this section we specialize to $D=5$ with $x^{5}\equiv y$.
We take as diffeomorphism the PBH diffeomorphism discussed earlier. The aim is to compute the on-shell action of the PBH mode $\tau$. From the last discussion we found all we need is the on-shell action in terms of the background, namely the solution before performing the PBH transformation, and a choice of time-like boundary surfaces, which we take to be:
\begin{align}\label{initialCOFFS}
y=y_{UV}, \ y=y_{IR}.
\end{align}
Under a generic PBH GCT this region transforms into:
\begin{gather}
-\infty < t, \ x <\infty, \ y^{\tau}_{IR}<y<y^{\tau}_{UV},
\end{gather}
with $y^\tau_{UV}$ and $y^\tau_{IR}$ given by the action (\ref{PBH}) on $y_{UV}$ and $y_{IR}$ respectively.
 In virtue of (\ref{VarAction}) we compute the transformed bulk action:
\begin{align}
S[y_{UV}]-S[y_{IR}]&=\int d^4x \int_{y_{IR}}^{y_{UV}}dy \sqrt{-g}L\\
 &\rightarrow \int d^4x \int_{y^{\tau}_{IR}}^{y^{\tau}_{UV}}dy \sqrt{-g}L=S[y^{\tau}_{UV}]-S[y^{\tau}_{IR}],\label{PBHAction}
\end{align}
where $y^{\tau}$ is given in (\ref{PBH}).
Given the near to boundary expansion of the bulk action for boundary metric $g^{(0)}=\eta$:
\[S_{div}=\int d^4x\left( \frac{1}{2 L_{UV}} \frac{1}{y_{UV}^2}+\frac{a_{UV}^{(2)}[\eta_{\mu \nu}]}{y
_{UV}}+a_{UV}^{(4)}[\eta_{\mu \nu}]\log (y_{UV})+...\right), \ \] with cut off surface at $y=y_{UV}$, we can then compute the leading terms in the PBH transformed effective action
by using (\ref{PBH}) and (\ref{PBHAction}):
\begin{multline}
\int d^4 x\frac{1}{y_{UV}^2}\rightarrow \int d^4 x \frac{1}{z_{UV}^2}-2\left(\frac{b^{(3)}(\partial\tau)^2+ b^{(5)}(\partial\tau)^4+ b^{(6)}\partial_{\mu}\tau \Box^{\mu
\nu}\tau\partial_\nu\tau}{z_{UV}^3}\right)+3\left( \frac{\left(b^{(3)}\right)^2(\partial
\tau)^4}{z_{UV}^4}\right)\\
\rightarrow \int d^4 x\left(\frac{1}{z_{UV}^2}+\frac{L_{UV}^2}{4}\frac{(\partial \tau)^2}{z_{UV}}+\frac{L^4_{UV}}{32}\left((\partial \tau)^4+2\partial_{\mu}\tau\partial_\nu\tau \Box^{\mu \nu}\tau
\right)...\right)\\
\rightarrow \int d^4 x\left(\frac{1}{z_{UV}^2}+\frac{L_{UV}^2}{4}\frac{(\partial \tau)^2}{z_{UV}}+\frac{L^4_{UV}}{32}\left( (\partial \tau)^4-4\Box \tau (\partial\tau)^2 \right)...\right).
\end{multline}
Similar contribution comes from the IR part of the primitive $S$. Should we demand IR smoothness of every background field, the static coefficients $a_{IR}^{(2)}[\eta_{\mu \nu}]$ and $a_{IR}^{(4)}[\eta_{\mu \nu}]$ will vanish automatically (See last paragraph in appendix \ref{sec:EoMApp}). So finally, we get the following form for the regularized bulk action:
\begin{gather}\label{bulkEffectiveAction}
S_{bulk}^{reg}=\int d^4 x  (\frac{e^{-2 \tau}}{2 L_{UV}y_{UV}^2}+\frac{L_{UV}e^{-\tau}}{8y_{UV}}(\partial \tau)^2+\frac{(L_{UV}^2 a_{UV}^{(2)}[\eta_{\mu \nu}]-L_{IR}^2 a_{IR}^{(2)}[\eta_{\mu \nu}])}{8}(\partial \tau)^2\nonumber\\ \ \ \hspace{5 cm}  +(a_{UV}^{(4)}[\eta_{\mu \nu}]-a_{IR}^{(4)}[\eta_{\mu \nu}])\tau+\frac{\Delta a}{8}\left( (\partial \tau)^4-4\Box \tau (\partial\tau)^2 \right))+\ldots,
\end{gather}
where $\Delta a= a_{UV}-a_{IR} $ with $a_{UV/IR}=\frac{L_{UV/IR}^3}{8}$. The $\ldots$ stand for logarithmic divergent terms that are going to be minimally subtracted. Notice that the gravitational Wess-Zumino term comes out with a universal coefficient $\Delta a$, independent of the interior properties of the flow geometry. Specific properties of the flow determine the normalization of the kinetic term and the Wess-Zumino term corresponding to the matter Weyl Anomaly. Next, we have to check whether this result still holds  after adding
the GH term and performing the holographic renormalization. So, from now on we restrict the discussion
 to the case of $\Delta=2$.
The finite Gibbons-Hawking contribution can be computed with the data given in appendix \ref{NTBExpansion}. One verifies that the contributions of
both boundaries  are  independent of derivatives of  $\tau$. The difference $S_{GH}|^{UV}_{IR}$ gives in fact a finite
contribution proportional to $\int d^4x \phi^{0} \tilde{\phi}^{(0)}$ which after a PBH tranformation reduces to a potential term for $\tau$. 

Notice that this term vanishes for a v.e.v. driven flow, so in this case no finite contribution at all arises. We will crosscheck this in the particular example studied in the next sections. In the case of a source driven flow, the finite contribution $\int d^4x \phi^{0} \tilde{\phi}^{(0)}$ give a potential term which is not Weyl invariant, as one can notice from the transformation properties (\ref{vev}).  In fact its infinitesimal Weyl transformation generates an anomalous variation proportional to the source square $\delta \tau \left(\frac{8\ L_{UV}^3}{3}(\tilde{\phi}^{(0)})^2\right)$. From the passive point of view, the GH term presents an anomaly contribution $\log(y_{UV})\left(\frac{8\ L_{UV}^3}{3}(\tilde{\phi}^{(0)})^2\right)$ that after the cut off redefinition originates a matter Wess-Zumino term $\int d^4 x \left(\frac{8\ L_{UV}^3}{3}(\tilde{\phi}^{(0)})^2\right) \tau $ (See equations (\ref{GHcoeff0}) and (\ref{GHcoeff1})).

Next, we analyze the counter-terms that are needed in order to renormalize UV divergencies.
Covariant counter-terms involve the boundary cosmological constant and curvatures for $g^{(0)}$ and the boundary values of the scalar field, namely v.e.v. and sources:
\begin{eqnarray}
\int d^{4}x\sqrt{\gamma } &=&\int d^{4}x\sqrt{g_{(0)}}(\frac{1}{y
_{UV}^{2}}+\frac{1}{y _{UV}}\frac{R}{12}+\frac{2}{3}\widetilde{\phi }%
_{(0)}^{2}+\frac{4}{3}\phi _{(0)}^{2}+...),\label{volumeCounterterm} \\
\int d^{4}x\sqrt{\gamma }R[\gamma ] &=&\int d^{4}x\sqrt{g_{(0)}}(\frac{R}{%
y_{UV}}+\frac{R^{2}}{12}), \\
\int d^{4}x\sqrt{\gamma }\Phi ^{2}(x,y _{UV}) &=&\int d^{4}x\sqrt{g_{(0)}}%
\left(\phi _{(0)}^{2}+\ldots\right),\label{finiteCounterterm}
\end{eqnarray}
where $\dots$ stand for logarithmic dependences that at the very end are going to be minimally substracted.
We take $g^{(0)}$ to be conformally flat and then use the Weyl transformation properties of the boundary invariants to compute
the Weyl factor dependence  of counter-terms.
The "volume" counter-term (\ref{volumeCounterterm}) is used to renormalize the infinite volume term of
an asymptotically $AdS_5$ space. One then needs to use the $R$ term to cancel the next to leading divergent term. In the process one remains with a finite potential contribution that even for a v.e.v.  driven flow gives a non vanishing energy-momentum trace contribution. The usual procedure \citep{deHaro, Bianchi1} is then to use the  finite covariant counter-term  (\ref{finiteCounterterm}) to demand conformal invariance in the renormalized theory, when the source is switched off.
The counter-term action satisfying this requirements is:
\[
S_{CT}=\int d^{4}x\sqrt{\gamma }\left(\frac{3}{2}-\frac{1}{8}R[\gamma
]-2 \Phi ^{2}\right)|_{UV}.
\]
 This action will provide an extra finite contribution to (\ref{bulkEffectiveAction}) proportional to:
\[\int d^4 x e^{-2 \tau} R^2[e^{-\tau}\eta]\sim \int d^4 x \left(\Box \tau - (\partial \tau)^2\right)^2 .\]
Finally the renormalized action takes the form:
\begin{eqnarray*}
S^{\Delta=2}_{ren} &=&S^{\Delta=2}_{reg}+S^{\Delta=2}_{GH}+S_{CT} \\
&=&\int d^4 x \left(\frac{16\ L_{UV}^{3}}{3}\tilde{\phi}_{(0)}^2 \tau + \frac{\Delta a}{8}\left((\partial \tau)^4-4 \Box \tau (\partial \tau)^2 \right)+\beta \left(\Box \tau -(\partial \tau)^2\right)^2+ O(1)+O\left(\partial^6 \right)\right).\\
\label{wesszumino}
\end{eqnarray*}
We should notice that no second derivative term, $(\partial\tau)^2$, is present in this particular case, just  as in the similar discussion of \citep{Sinha2}.
However, there is a source of higher derivative terms: due to the fact that the PBH 
diffeomorphism is  singular in the IR, and in fact,
the higher orders in derivatives come with the higher order IR singularities. So, the higher derivative terms  are counted by powers of the IR cut off.
We do not address here  the issue of renormalizing these terms. The main idea here was to show the presence of a Wess-Zumino term compensating the anomaly difference between fixed points. 
The term $O(1)$ stands for possible finite contributions (4D cosmological constants) in the static on shell action plus GH term and CT. As for the GH term this contributions vanish for v.e.v. driven flows.
\newpage

\section{RG Flow in $N=4$ 3D Gauged Supergravity}
In this section we consider a particular, explicit and analytic example of a  Holographic RG flow in 3D
gauged supergravity. The reason to analyze this particular example is twofold: first,  it is relatively simple
and  analytic, and,  second, it is completely smooth, even in the infrared region. Indeed smoothness will be our guiding
principle in deforming the background geometry in the way we will detail in this section. We will
promote some integration constants (moduli) present in the flow solution
to space-time dependent fields and identify among them the one which corresponds to
a specific field in the boundary CFT.
To get still a solution of the equations of motion
we will have to change the background to take into account the back reaction of space-time derivatives
acting on the moduli fields. This will be done in a perturbative expansion in the number
of space-time derivatives.
The starting point is one of the  explicit examples  of  RG flows  studied in \citep{Parynha},
where domain wall solutions in $N=4$ 3D gauged supergravity were found. These solutions
are obtained by analyzing first order BPS conditions and respect 1/2 of the bulk supersymmetry.
They describe holographic RG flows between $(4,0)$ dual SCFT's. It turns out that
the solution we will be considering admits a consistent lift to 6D supergravity, which
will be reviewed and used in the next section. In this section the analysis will
be purely three-dimensional.


We start by writing the action and equations of  motion for the three dimensional theory at hand. In this case the spectrum reduces to the metric $g$, and a pair of scalars $A$ and $\phi$, which are left over after  truncating the original scalar manifold. The action is:

\begin{align}\label{3DAction}
S^{bulk}_{scalars} =\int d^{3}x \sqrt{-g} \left(-\frac{R}{4}-\frac{3}{4}\frac{( \partial A )^2}{ \left(1-A^2\right)^2}-\frac{1}{4}( \partial \phi )^2
-V(A,\phi )\right),
\end{align}
with potential for the scalar fields given by:
\begin{align}\label{potential}
V&=\frac{1}{2} e^{-4 \phi} \left(\frac{2 e^{2 \phi } \left(A^2 \left(g_2 A \left(g_2 A \left(A^2-3\right)+4
   g_1\right)-3 g_1^2\right)+g_1^2\right)}{\left(A^2-1\right)^3}+4 c_1^2\right).
\end{align}
The corresponding set of equations of motions is then given by:

\begin{eqnarray}\label{EoM}
\frac 12\Box \phi -\partial _\phi V(A,\phi ) &=0,\label{phiEq} \\
\frac 32\frac 1{\sqrt{-g}}\partial _\mu (\frac 1{(1-A^2)^2}g^{\mu \nu
}\partial _\nu A)-\partial _AV(A,\phi ) &=0 ,\label{AEq}\\
-\frac 14R_{\mu \nu }-\frac 34\frac{\partial _\mu A\partial _\nu A}{(1-A^2)^2%
}-\frac 14\partial _\mu \phi \partial _\nu \phi -g_{\mu \nu }V(A,\phi ) &=0. \label{EinstEq}
\end{eqnarray}

\subsection{The domain wall solution and its moduli}
In this subsection we  review the domain wall solution describing  the RG flow
on the dual CFT and identify its moduli.
 Let us choose coordinates $x^\nu=t,x,r$ and
the 2D $(t,x)$-Poincar\'{e} invariant domain wall ansatz for the line element:
\begin{align}
ds^2 =dr^2 + e^{2f(r)}\eta_{\mu\nu}dx^{\mu}dx^{\nu},
\end{align}
and the scalar field profiles $A_B(r)$ and $\phi_B(r)$.

 The equations of motion reduce then to the following set:
\begin{align}\label{EoMBack}
f' \phi_B'+\frac{\phi_B''}{2}-\partial_{\phi_B} V  &=0,\\
3 A_B''+6 A_B' f'+6 \frac{A_B A_B'^2}{(1- A_B^2)}-2
   \left(1-A_B^2\right)^2 \partial_{A_B} V&=0,\\
 \left(2 f''+2 f'^2+\phi_B'^2+\frac{3 A_B'^2}{\left(A_B^2-1\right)^2}+4 V\right)&=0,
 \end{align}
where the primes denote derivative with respect to $r$.
It is then straightforward to show that the following field configuration:
\begin{gather}\label{backSol}
e^{\phi_B(r)}=\frac{2 c_1 \left(g_2^2-\frac{g_1^2 \rho^2}{\left(\rho+y(r)\right)^2}\right)}{g_1 g_2^2 \sqrt{1-A_B(r)^2}}, \\
A_B(r)=\frac{g_1}{g_2}\frac{\rho}{ \left(\rho+y(r)\right)}, \
e^{2f(r)}=\frac{1}{2}e^{2 s_p}y(r)\left(\frac{g_2^2 \left(\rho+y(r)\right)-g_1^2\rho}{ \left(\rho+y(r)\right)}\right)^2,
\end{gather}
with $y(r)=e^{2 g_1 F(r)}$ is the most general solution of (\ref{EoMBack}),(\ref{AEq}), (\ref{phiEq}), provided:
\begin{align}
F'(r)=\frac{g_1 g_2^2 \left(\rho+y(r)\right)^2}{2 c_1 \left(g_2^2 \left(\rho+y(r)\right)^2-g_1^2\rho^2\right)}.
\end{align}
We can solve this equation explicitly for r(F):
\begin{align}\label{rToF}
r(F)&=\frac{2 c_1 \left(F \left(g_2^2-g_1^2\right)-\frac{g_1}{2 \left(e^{2 F g_1}+\rho\right)}+\frac{1}{2} g_1 \log \left(e^{2 F
   g_1}+\rho\right)\right)}{g_1 g_2^2}+\tau.
\end{align}
Notice the presence of three moduli $\tau$, $s_p$, $\rho$. The first one corresponds to  a freedom in shifting the radial
coordinate by a constant amount $\tau$, $r \rightarrow r+\tau $. This mode is a PBH rigid diffeomorphism in the domain wall coordinates. As mentioned a rigid PBH in domain wall coordinates becomes a warped one in the Fefferman-Graham coordinates. 
The second modulus $s_p$ can be identified with a rigid conformal
transformation in the boundary coordinates $(t,x)$.
The third modulus $\rho$ is an internal mode respecting the boundary conditions for the metric in both UV and IR limits
but changing the scalar modes and It corresponds to a normalizable zero mode.
In the next section we will see this mode is basically the instanton size modulus in the 6D description
of the RG flow. But can be also thought of as a linear combination of a PBH and $s_p$ mode.
In order to have a flavor of the properties of the flow geometry it is useful to make a change of coordinates, from $(t,x,r)$ to $(t,x,y)$ with $y=e^{2 g_1 F(r)}$.
In this coordinates the metric becomes:
\begin{align}\label{metricY}
ds^2&=\frac{\left(g_2^2 (y+\rho)^2-g_1^2 \rho^2 \right)}{2 (y+\rho)^2} \left(\frac{2 c_1^2 \left(g_2^2 (y+\rho)^2-g_1^2 \rho^2 \right)}{g_1^4 g_2^4 y^2(y+\rho)^2}dy^2+e^{2 s_p} y \left(dx^2-dt^2\right)\right).
\end{align}
This geometry approaches $AdS_3$  in both the UV($y \rightarrow \infty$) and the IR ($y \rightarrow 0$) limits, with corresponding radii:
\begin{align}
\frac{L^{2}_{IR}}{4}=\frac{ c_1^2 \left(g_1^2-g_2^2\right){}^2}{g_1^4 g_2^4 } \text{ and } \frac{L^{2}_{UV}}{4}=\frac{c_1^2}{g_1^4}.
\end{align}
These radii determine the central charges of the (4,0) CFT's at the fixed points, through the expression $c=3 L/2G_N$,
$G_N$ being the 3D Newton's constant \footnote{In our conventions $G_N$=4.}.
Additionally the limit: \begin{align}g_2\rightarrow \infty  \text{ with } g_1 \text{ fixed},  \label{AdS3Limit}\end{align} recovers $AdS_3$  space with radius $L$ given by $\frac{L^{2}}{4}=\frac{
c_1^2}{g_1^4}$. An additional transformation in the boundary metric is needed to keep it finite in the limit,
$\eta\rightarrow \frac{2}{g_2^2}\eta$.
The scalar fields go in the UV and IR to different fixed points (extrema) of the potential
$V(A,\phi)$. In particular, in the UV, $A\rightarrow 0$ and $\phi\rightarrow \log\left(\frac{2c_1}{g_1}\right)$.
Expanding the potential ($\ref{potential}$) around the extremum we find out the masses of the bulk fields $A(r)$, $\phi(r)$
at the UV fixed point:
\begin{align}
m_{A}^2=0 \,\text{ and } m_{\phi}^2=\frac{h^2 g_1^4}{c_1^2}=\frac{8}{L_{UV}^2}.
\end{align}
The allowed conformal dimension of the corresponding dual boundary operators are:
\begin{align}
{\Delta_{A}}_+=2,\, {\Delta_{A}}_-=0\,\text{  and  } \Delta_{\phi}=4,
\end{align}
respectively. By
looking at (\ref{backSol}) we can read off their asymptotic expansions near the UV boundary ($y\rightarrow \infty$):
\begin{align}\label{vevs}
\delta A(y) \sim\frac{g_1}{g_2}\frac{\rho}{y} \text{ and }
\delta \phi(y)\sim -\frac{g_1^2}{2 g_2^2}\frac{\rho^2}{y^2}.
\end{align}
These are "normalizable" excitations, and in the standard quantization, which adopts $\Delta=\Delta_+$,
they would correspond  to a vacuum state in the dual CFT,
where the  dual operators ${\cal O}_A$ and ${\cal O}_\phi$ acquire a v.e.v..  
This clashes with the fact that in $D=2$ we cannot have spontaneous breaking
of conformal invariance \footnote{On the other hand, the "alternate" quantization \citep{kw} would interpret this background as a source term for the $\Delta_-=0$ operator ${\cal O}_A$.
However, this interpretation clashes with the standard axioms of 2D CFT.}. Notice that the problem arises also in the well known case of the D1-D5 system
in IIB, when one deforms the $AdS_3\times S^3$ background  by going to
multi-center geometries. 
Most probably this is a feature of the supergravity approximation, or dually, of the leading large-$N$ expansion
on the CFT side. It would interesting to see how the picture is modified in going beyond the supergravity
approximation, as discussed, in a different context, in \citep{hartnoll}. 



At the IR,
\begin{align}
\delta A(y) \sim-\frac{g_1 }{g_2}\frac{y}{\rho} \text{ and }
\delta \phi(y)\sim -\frac{g_1^2}{g_1^2-g_2^2}\frac{y}{\rho} .
\end{align}
In particular,  the background is completely smooth. Now we notice a property of the metric (\ref{metricY}) : the UV/IR AdS limits of the geometry are independent of $\rho$, and, as mentioned earlier,  this modulus corresponds to  a normalizable zero mode.


It is instructive to look at how can be represented a PBH diffeomorphism zero-mode of the form $y\rightarrow e^{2\sigma_{PBH}} y$ in terms of the moduli appearing
in the background geometry: it amounts to take the combined set of transformations $\rho \rightarrow \rho e^{2\sigma_{PBH}}$ and $s_p\rightarrow sp+\sigma_{PBH}$. Conversely, the $\rho$ modulus can be thought of as a combination of a PBH mode mentioned before plus a suitable choice of $s_p$ such that the boundary metric remains unchanged.
We should stress that the PBH zero-modes $\tau$ and $\sigma_{PBH}$ aren't precisely the same. The difference will come about in the next subsection. But we can already say that there is a choice of $\tau$ and $s_p$ for fixed $\rho=1$ that preserves normalizability.
We can explore then two possibilities, either we analyze the combined pair of moduli $(\tau, \ s_p)_{\rho=1}$ or the single modulus $\rho$.
In the next subsection we analyze both cases. We will also check the geometrical procedure discussed in section \S 1.

\subsection{Fluctuations Analysis}
In this subsection we are going to analyze a deformation of the background geometry which
arises when one gives a non trivial $(t,x)$ dependence to some of the moduli introduced
in the previous subsection. Specifically, we will promote the integration constants  $s_p$ and $\tau$
to functions of $t$ and $x$, $s_p(t,x)$ and $\tau(t,x)$. In doing so, of course,
we have to take into account the back reaction due to the $(t,x)$ derivatives acting these fields. The equations of motion
will involve therefore inhomogeneous terms containing derivatives of $s_p(t,x)$ and $\tau(t,x)$.  We will work
in a perturbative expansion in the number of $t$ and $x$ derivatives. For that purpose it is convenient to introduce
a counting parameter $q$, whose powers count  the number of $t,x$ derivatives.
As for the metric,
we keep the axial gauge condition and therefore start with the expression:
\begin{align}\label{metric2}
ds^2=dr^2+(e^{2 f}\eta_{\mu \nu}+q^2 g_{\mu \nu}^{(2)})dx^{\mu}dx^{\nu},
\end{align}
where $x_{0}=t$ and $x_1 =x$, and $\mu$, $\nu=0,1$. 

For the background deformations, at second order in $(t,x)$ derivatives, we adopt the following ansatz 
for the scalar fields:
\begin{eqnarray}
A&=&A_B+q^2 A^{(2)}\nonumber\\
\phi&=&\phi_B+q^2 \phi^{(2)}
\end{eqnarray}
whereas for the metric components:
\begin{equation}
g_{tt}^{(2)}=-e^{2f}(g^{(2)}+ T),\, g_{xx}^{(2)}=e^{2f} ( g^{(2)}-T), 
\end{equation}
and we redefine $g_{tx}^{(2)}\rightarrow e^{2f} g_{tx}^{(2)}$.
The homogeneous part of the
equations of motion will involve an ordinary linear differential operator in the $r$ variable acting on the fluctuations and this will
be sourced by an inhomogeneous term involving two $t,x$ derivatives acting on $s_p$ and $\tau$, which represents
the moduli back reaction to the original background.
Now we have five unknown functions  and eight equations, 
(\ref{phiEq}), (\ref{AEq}), (\ref{EinstEq}), so that we need to reduce the number  of independent equations. It is a long but straightforward procedure to find out the general
 solutions to the system. We are going to sketch  the procedure we followed to solve them. Details are given in appendices. Specifically the equations of motions at order $q^2$ are given in appendix \ref{EqFluctuations}.

A change of coordinates is useful to render the system of partial differential equations simpler. We perform a  change from the domain wall coordinates $(t,x,r)$ to the Poincar\'{e} like coordinates $(t,x,y)$ already introduced in the previous subsections:
\begin{align}
y=e^{2 g_1 F(\tau(t,x),r)},
\end{align}
where,
\begin{align}
\partial_r F-\frac{g_1 g_2^2 \left(e^{2 g_1 F}+1\right)^2}{2 c_1 \left(g_2^2 \left(e^{2 g_1
F}+1\right)^2-g_1^2\right)}=0.
\end{align}
Notice that if we are using a
non fluctuating cut off surface $r=r_{UV}$ in the original coordinates, in the new coordinates the same surface will be fluctuating at a pace
dictated by $\tau(t,x)$. We can however use a different choice of coordinates:
\begin{align}
\widetilde{y}=e^{2 g_1 F(0,r)}.
\end{align}
It is then easy to show based on (\ref{rToF}), that cut offs shapes in the $y$-system and $\widetilde{y}$-system are related as follows:
\begin{align}
y_{UV} \rightarrow  e^{\frac{g_1^2  }{c_1 }\tau} \widetilde{y}_{UV}, \
y_{IR} \rightarrow e^{\frac{g_1^2 g_2^2  }{c_1 \left(g_2^2-g_1^2\right)}\tau} \widetilde{y}_{IR}.
\end{align}


The set of equations, (\ref{phiEq}), (\ref{AEq}), (\ref{EinstEq}) provides a system of second order differential equations for the fluctuations
in terms of the inhomogeneities produced by derivatives acting on  $s_p(t,x)$ and $\tau(t,x)$.
We are going to denote the five Einstein equations (\ref{EinstEq}) by $(t,t)$, $(x,x)$, $(t,x)$, $(r,r)$, $(t,r)$, $(x,r)$,
with obvious meaning.
Equations $(t,t)$, $(x,x)$ and $(r,r)$ form a set of second order equations in the $\eta$-trace part of the metric parametrized by $g^{(2)}(t,x,r)$
and the traceless part parametrized by $T(t,x,r)$, together with the scalar fluctuations, which only appear up to first order in radial derivatives. It turns out that the
combination $(t,t)-(x,x)$ gives an equation for the trace part and scalar  fluctuations, but the traceless part decouples in the
combination $(t,t)+(x,x)$. Namely it gives the equation:
\begin{equation}
y\partial_y^2 T+2\partial_y T+\frac{2e^{-2s_p} 2g_1^2(g_1^2+3g_2^2(1+y^2))\left((\partial_t^2\tau)^2+(\partial_x^2\tau)^2\right)}
{(g_1^2-g_2^2(y^2+1))^3}=0,
\label{Tequation}
\end{equation}
whose general solution is:
\begin{align}
T=C_3(t,x)-\frac{1}{y}C_2(t,x)+\frac{g_1^2}{g_2^2 y \left(g_2^2 (y+1)^2-g_1^2\right)}e^{-2 s_p}  \left((\partial_t \tau)^2+(\partial_x \tau)^2\right),
\end{align}
where $C_3$ and $C_2$ are integration constants promoted to be arbitrary functions of $t$ and $x$.
Let's focus then on the set of equations $(t,t)-(x,x)$ and $(r,r)$. This is a coupled system for the trace part and the
scalars which can be solved in many different ways, here we present one. First of all $(r,r)$ can be integrated to get:
\begin{align}\label{rrIntegrated}
\partial_y g^{(2)}=R^{(1)}_{\partial_y g^{(2)}} A^{(2)}+R^{(2)}_{\partial_y g^{(2)}} \phi^{(2)}+\frac{1}{y^2} C_{5},
\end{align}
where,
\begin{align}
R^{(1)}_{\partial_y g^{(2)}}=-\frac{6 g_1 g_2^3 (y+1)^2}{\left(g_1^2-g_2^2 (y+1)^2\right){}^2}, \ R^{(2)}_{\partial_y g^{(2)}}=
-\frac{2 g_1^2}{(y+1) \left(g_2^2 (y+1)^2-g_1^2\right)},
\end{align}
with an integration constant $C_5$.
Then, one can notice that Eq. $(t,t)-(x,x)$ only contains derivatives of the trace part of the metric fluctuations, so we can use (\ref{rrIntegrated})
and its derivative to eliminate this function. The remaining equation will contain the scalar fluctuations up to first order in "radial"
derivatives:
\begin{align}\label{phi2list}
\partial_y \phi ^{(2)}&=R^{(1)}_{\partial_y \phi ^{(2)}} \partial_y A^{(2)}+R^{(2)}_{\partial_y \phi ^{(2)}}\phi ^{(2)}+R^{(3)}_{\partial_y
\phi
^{(2)}}A^{(2)}\nonumber\\& \qquad+
R^{(4)}_{\partial_y \phi ^{(2)}}(C_5-\frac{2 c_1}{g_1^2 g_2^2}\square \tau)+R^{(5)}_{\partial_y \phi ^{(2)}}(\partial \tau)^2
+R^{(6)}_{\partial_y \phi ^{(2)}}e^{-2s_p}\square s_p. \end{align}

Under the conditions already found the remaining equations (\ref{phiEq}), (\ref{AEq}) reduce to the final algebraic equation for
$\phi^{(2)}$ in terms of y-derivatives of $A^{(2)}$ up to second order. By solving it and plugging the result in (\ref{phi2list}) we obtain the
third order differential equation:
\begin{align}\label{ThirdOrderEquation}
\partial^{(3)}_y A^{(2)}+R^{(2)}_{A^{(2)}}\partial^{2}_y A^{(2)}+
R^{(1)}_{A^{(2)}}\partial_y A^{(2)}+R^{(0)}_{A^{(2)}}A^{(2)}=e^{-2 s_p}F,
\end{align}
where the inhomogeneous part takes the form:
\begin{align}
F=F^{(1)}C_5+F^{(2)}\square s_p +F^{(3)}\square \tau +F^{(4)}(\partial \tau)^2.
\end{align}
The $R^{(i)}_{A^{(2)}}$ and $F^{(i)}$ are rational functions in the radial coordinate $y$ ( They are given in the appendix \ref{rationalFunctions}). We solve this equation by Green's function method (See appendix \ref{GreenF}).


The $(t,x)$ equation:
\begin{align}
\partial^{2}_y g^{(2)}_{t x}=-\frac{2}{y} \partial_y g^{(2)}_{t x} + \frac{4 g_1^2 e^{-2 s_p} \left(3 g_2^2 (y+1)^2+g_1^2\right)}{y \left(g_2^2
(y+1)^2-g_1^2\right){}^3} e^{-2 s_p} \partial_t \tau \partial_x \tau,
\end{align}
can be solved to get:
\begin{align}
g^{(2)}_{t x}=-\frac{C_6(t,x)}{y}+C_7(t,x)-\frac{2 g_1^2}{g_2^2 y \left(g_2^2 (y+1)^2-g_1^2\right) }e^{-2 s_p}\partial_t \tau \partial_x \tau.
\end{align}


As for the mixed equations, $(t,r)$ and $(x,r)$, they involve odd number of $(t,x)$ derivatives
and one needs to go to third order, were  in fact they reduce
to differential constraints for the integration constants $C_2$, $C_{5}$ and $C_6$ sourced by second derivatives of the moduli $\tau$ and $s_p$.
Before solving for these constraint equations it is convenient to analyze the constraints that IR regularity imposes on the modulus $C_5$.

At this point we should comment about an important issue. We have nine integration functions $C_i(t,x)$ and our general
on shell fluctuations develop  generically infrared singularities and/or UV non-normalizabilty, in the latter case representing
source terms on the dual CFT.
We have two ways to deal with  possible IR divergencies in our deformed background geometry:
we could allow infrared singularities of the geometry and put a cut off at the IR side, or demand IR-smoothness. This last option will
spoil full normalizabilty of all fluctuations, as we will see. This is
something perhaps  we could allow  because at $q^0$ order the modulus which could be associated to the "dilaton"
is still a normalizable bulk mode. The first
option will guarantee full normalizability to order $q^2$, but will require the presence of an IR Gibbons Hawking (GH) term (\ref{IRGH}).
In any case we will see that the GH term will give no contribution to the boundary effective action of the moduli.
In this paper we take the first point of view and demand full smoothness of the deformed geometry.
By demanding regularity in the IR side for our spectrum of matter fluctuations $A^{(2)}$ and $\phi^{(2)}$ we get the following set of relations
for the integration functions:

\begin{align}\label{C5}
C_5(t,x)&=-\frac{2 c_1 }{g_1^2 g_2^2}e^{-2 s_p} \Box \tau +\frac{4 c_1^2 \left(g_1^2-g_2^2\right) }{g_1^4 g_2^4}e^{-2 s_p} \Box s_p, \\
C_{10}(t,x)&=\frac{9 g_1^5 }{4 g_2^9} e^{-2 s_p}(\partial \tau)^2-\frac{c_1^2 \left(9 g_1^4-17 g_2^2 g_1^2+8 g_2^4\right)}{2 g_1 g_2^{11}} e^{-2
s_p} \Box s_p.
\end{align}
At this point we could solve the $(t,r)$ and $(x,r)$ fluctuation equations for the moduli:
\begin{align}
e^{2 s_p}C_2(t,x)&=\frac{4 c_1^2 \left( g_1^2-g_2^2\right)}{g_1^4 g_2^4}\left((\partial_{t} s_p)^2-\partial^{2}_{t} s_p\right) -\frac{4 c_1}{g_1^2
g_2^2}
\partial_{t} s_p \partial_{t} \tau -\frac{1}{g_2^2} (\partial_{t} \tau)^2+\frac{2 c_1}{ g_1^2 g_2^2}\partial^{2}_{t} \tau+\left(\partial_t
\rightarrow \partial_x \right),\\
e^{2 s_p}C_6(t,x)&=-\frac{8 c_1^2 \left( g_1^2-g_2^2\right)}{g_1^4 g_2^4}(\partial_{t} s_p\partial_{x} s_p-\partial^2_{tx} s_p) +\frac{4 c_1}{g_1^2
g_2^2} (\partial_{t} s_p \partial_{x} \tau+\partial_{x} s_p \partial_{t} \tau-\partial^2_{tx} \tau)+\frac{2}{g_2^2}(\partial_{t} \tau
\partial_{x} \tau).
\end{align}
  According to the AdS/CFT dictionary, a state in the boundary CFT should correspond
to a normalizable bulk mode, whereas non normalizable modes correspond to source deformations of the CFT.
In our case, the UV boundary metric in the Fefferman-Graham gauge looks like $e^{2s_p+\frac{g_1^2}{c_1} \tau} \eta$. So, assuming the "standard" quantization, if we don't want to turn on sources for
the trace of the boundary energy momentum tensor we need to take:
\begin{align} \label{normalizability}
\tau = -\frac{2 c_1}{g_1^2} s_p.
\end{align}
This is not the case in the IR boundary where the induced metric picks up a shifting factor that we cant avoid by staying in the axial gauge
($g_{rr}=1$). By requiring not to turn on sources, even at second order in the derivative expansion for other
components of the UV boundary CFT stress tensor, we see that:
\begin{align}
C_3(t,x)=0, \ C_4(t,x)=0 , \ C_7(t,x)=0.
\end{align}
At this point of the nine integration constants at our disposal, after requiring regularity and normalizability of the metric fluctuations, two are left over,  $C_8$ and
$C_9$. Together with $\tau$ they determine the CFT sources inside the matter fluctuations $\phi^{(2)}$ and $A^{(2)}$. This remaining freedom can be used
just to require normalizability of either $\phi^{(2)}$ or $A^{(2)}$, but not both of them. From here onwards we choose to make $\phi_{(2)}$ normalizable but for our purposes the two choices are equivalent. Finally we get:
\begin{multline}
C_9(t,x)=4 C_8(t,x)+\frac{\left(-3 g_1^7+13 g_2^2 g_1^5-4 g_2^4 g_1^3\right) }{g_2^7}e^{-2 s_p}(\partial \tau)^2\\+\frac{2 c_1^2 \left(27 g_1^8-144 g_2^2 g_1^6+139 g_2^4 g_1^4+23 g_2^6 g_1^2-12 g_2^8\right)}{9 g_1^3 g_2^9} e^{-2 s_p} \Box s_p.
\end{multline}
This choice turns on a source for the CFT operator dual to $A$. Indeed the UV expansion for $A$-fluctuation reads:
\begin{align}\label{sourceA}
A^{(2)}\sim-\frac{2 c_1^2}{3 g_1^3 g_2^3} e^{-2 s_p} \left(\Box s_p\right).
\end{align}
To summarize, requiring IR regularity forces us to turn on a source term for one of the scalar fields.
Notice that under the condition (\ref{normalizability}) the traceless and off-diagonal modes $T$ and $g^{(2)}$ are IR divergent. They go
as $\frac{1}{y}$ in the IR limit. Nevertheless the IR limit of the metric is not divergent because of the extra warp factor, which is proportional to $y$. Notice that The AdS IR limit is in fact broken by $q^2$ order fluctuations, as already argued in section \S 1.

\subsection{Evaluating the on-shell Action}
The regularized boundary Lagrangian coming from the bulk part is obtained
by performing the integral over the radial coordinate with IR and UV cut-offs $y_{IR}$ , $y_{UV}$ respectively:
\begin{align}
L^{bulk}_{2D}=\int^{y_{UV}}_{y_{IR}} dy \textsl{L}_{3D}.
\end{align}
First we present the result for the presence of both the moduli $s_p$ and $\tau$. We can write down the 3D lagrangian as:
\begin{multline}
L_{3D}=l^{(0)}+ l^{(1)}(\partial \tau)^2 +l^{(2)}\Box \tau +l^{(3)} \Box sp +\partial_y \left(l^{(4)} \partial_y g^{(2)}_{tt}
+l^{(5)}g^{(2)}_{tt}+l^{(6)}A^{(2)}+l^{(7)} \phi^{(2)}) \right).
\end{multline}
After integration and evaluation at the cut off surfaces we arrive to a boundary regularized action:
\begin{multline}\label{LagOnshell}
\int dtdx L^{bulk}_{2D}=\int dtdx ( \frac{g_1^2 g_2^2}{8 c_1}e^{2 s_p(t,x)}\left[y\right]^{UV}_{IR}
-\frac{g_1^4}{8 c_1 \left(g_1^2-g_2^2\right)}(\partial \tau)^2+\left(\frac{1}{4} \Box \tau+\frac{ c_1}{2 g_1^2}\Box s_p \right) \log
y_{UV}\\-\left(\frac{1}{4}\Box \tau+\frac{c_1(g_2^2-g_1^2)}{2 g_1^2 g_2^2}\Box s_p \right) \log y_{IR}+...+\left[L_{hom}\right]^{UV}_{IR}),
\end{multline}
where the $...$ stand for infinitesimal contributions and a total derivative term \[-\frac{c_1}{2 g_2^2} \Box s_p
+\log(1-\frac{g^2_2}{g^2_1})\Box \tau,\] which is irrelevant for the discussion. Notice that the logarithmic
divergent part is a total derivative, as it should be. Moreover the coefficient in front of It is proportional to the difference of central charges at the UV and
IR fixed points. The contribution of the homogeneous part of the solutions to the onshell bulk action can be written as:
\begin{align}\label{homogeneous1}
L_{hom}=\left(l^{(4)} \partial_y g^{(2)}_{tt} +l^{(5)}g^{(2)}_{tt}+l^{(6)}A^{(2)}+l^{(7)} \phi^{(2)}\right).
\end{align}
As we will show in a while, this contribution does not affect the finite value of the moduli $\tau$ and $s_p$ effective action at all! In next section we will see this will not be the case if we work in Fefferman-Graham gauge since the beginning. In that case, the solution of homogeneous equations do affect the final result but upon regularity conditions the contributions are total derivatives of the moduli and hence irrelevant. The explanation in this mismatch comes from the fact the coordinate transformation from one gauge to the other is singular at $q^2$ order.
After using (\ref{rrIntegrated}) on (\ref{homogeneous1}) we get:
\begin{align}\label{homogeneous2}
L_{hom}=\left(  l^{(5)}g^{(2)}_{tt}+\left(l^{(6)}+l^{(4)}\times R^{(1)}_{\partial_y g_{tt}^{(2)}}\right)A^{(2)}+\left(l^{(7)}+l^{(4)} \times
R^{(2)}_{\partial_y g_{tt}^{(2)}}\right) \phi^{(2)}\right).
\end{align}
Now, we asymptotically expand $L_{hom}$. For this we need to use the most general form of the solutions to $g^{(2)}_{tt}$,
$A^{(2)}$
and $\phi^{(2)}$. After a straightforward computation one gets:
\begin{align}
L_{hom}&\xrightarrow[y\rightarrow y_{UV}]{} \frac{g_1^2 g_2^2 }{8 c_1}e^{2 s_p}C_5(t,x) +O\left(\frac{1}{y_{UV}}\right), \\
L_{hom}&\xrightarrow[y\rightarrow y_{IR}]{} \frac{g_1^2 g_2^2 }{8 c_1}e^{2 s_p}C_5(t,x) +O\left(y_{IR}\right).
\end{align}
The only integration constant entering the boundary data is given by $C_5(t,x)$. However $\left[L_{hom}\right]^{y_{UV}}_{y_{IR}}$ vanishes,
and the boundary effective action for the moduli $s_p$ and $\tau$ coming from the bulk action is independent of all the integration constants, namely, any particular solution of the inhomogeneous system of differential equations gives the same final result, so far. We say so far, because still we have not commented about the GH and CT contributions. This is an interesting outcome, since the result  holds independently of the IR regularity and normalizability conditions imposed
on the fluctuations discussed earlier. The GH term will not affect this observation, but the CT contribution does it. In any case, we choose integration constants in order to satisfy our cardinal principle: IR regularity.

\subsection{Gibbons-Hawking contribution}
Let us discuss now the GH contribution: in the domain wall coordinates, (\ref{metric}), it reads:
\begin{align}
\frac{1}{2}\int dtdx L^{GH}_{2D}= \frac{1}{2}\int dtdx \sqrt{g^{rr}} \partial_r \left( \sqrt{g^{rr}} \sqrt{-\det{g}} \right)|_{boundary},
\end{align}
where so far $g_{rr}=1$, but for later purposes it is convenient to write the most general form above. In the $(t,x,y)$ coordinates and after
using
(\ref{rrIntegrated}) it is simple to show that:
\begin{align}
L^{GH}_{2D}=\left(-\frac{g_1^2 g_2^2  y \left(g_2^2 (y+1)^3+g_1^2 (y-1)\right)}{4 c_1 (y+1) \left(g_2^2 (y+1)^2-g_1^2\right)}e^{2 s_p}-2
L_{hom}\right) |_{boundary},
\end{align}
The UV and IR asymptotic expansions are thence given by:
\begin{align}\label{UVGH}
L^{GH}_{2D} &\xrightarrow[y\rightarrow y_{UV}]{}-\frac{g_1^2 g_2^2}{4 c_1}e^{2 s_p} y_{UV}
+O\left(\frac{1}{y_{UV}}\right),\\\label{IRGH}
L^{GH}_{2D} &\xrightarrow[y\rightarrow y_{IR}]{}\frac{g_1^2 g_2^2}{4 c_1}e^{2 s_p} y_{IR}
+O\left(y_{UV}^2\right).
\end{align}
Even though we are not taking the approach of cutting off the geometry in the IR side, we present the IR behaviour of GH term just for
completeness of analysis. Notice there is not finite contribution coming from them and again the independence on integration constants mentioned previously. 

\paragraph{Regularized Action}
At this point we can write down the regularized Lagrangian for the "normalizable" modulus $s_p$. We first make the change to the
Fefferman-Graham gauge at $q^0$ order, $y\rightarrow \tilde{y}$, make use of the normalizability condition (\ref{normalizability}) and the final result becomes:
\begin{align}
S^{2D}_{reg}= \int dt dx\left( \frac{g_1^2 g_2^2}{8 c_1}\tilde{y}_{UV}-\frac{c_1}{2 g_2^2}\Box s_p \log
\tilde{y}_{IR}+\frac{1}{2}\frac{c_1}{(g_2^2-g_1^2)} (\partial s_p)^2 +...\right),
\end{align}
where the $...$ stand for subleading contributions in term of the cutoffs and finite total derivative terms. Notice that there is no
logarithmic divergence at the UV cutoff. This is because this modulus is not affecting the UV boundary
metric. On the other hand the IR side does have a logarithmic divergent factor, which however is a total derivative.

Now, we discuss  possible contributions coming from covariant counterterms.
Let us start by gravitational countertems. In the asymptotically AdS$_3$
geometries the leading divergence in the on-shell action is renormalized by using the covariant term \[ \int d^2 x \sqrt{-\det{\gamma}}|_{bdry}
=
\int d^2 x  \frac{g_2^2}{2} \tilde{y}_{UV}+ \frac{c_1}{g_1^2}\left( \frac{2 c_1}{g_1^2} \Box s_p +\Box \tau
\right)+O(\frac{1}{\tilde{y}_{UV}}).
\]
Other possible counterterms are:
\begin{gather}\label{CTUV}
\int dt dx \sqrt{-\gamma} R^{(2D)}[\gamma]|_{bdry} =\frac{1}{c_1}\left( \frac{2 c_1}{g_1^2} \Box s_p +\Box \tau
\right)+O(\frac{1}{\tilde{y}_{UV}}) , \\ \label{CTUV2}
\ \int dt dx \sqrt{-\gamma} (\delta A)^2|_{bdry}=-\frac{2 c_1^2}{3 g_1^2 g_2^2} \Box s_p, \ \int dt dx \sqrt{-\gamma} (\delta \phi)^2|_{bdry}=
O\left( \frac{1}{\tilde{y}_{UV}^3} \right),
\end{gather}
where $\delta A$, $\delta \phi$ denote the fluctuations around the UV stationary point of the potential.
Notice that after imposing the normalizability condition (\ref{normalizability}) the finite contributions of this counterterm disappear except
for the $\delta A$ fluctuation which is a total derivative contribution. The remaining IR logarithmic divergence is minimally subtracted.
Finally the renormalized action takes the form:
\begin{align}\label{S2Drenormalized}
S^{2D}_{ren}= \int dt dx\left(\frac{1}{2}\frac{c_1}{(g_2^2-g_1^2)} (\partial s_p)^2+O\left(\partial^{4}\right)\right).
\end{align}
The coefficient in front of this action is not the difference of central charges of the UV/IR fixed points. Although we can always rescale the field, this mismatch is unpleasant, because a rigid shifting in the spurion mode $\tau$ (not on $s_p$) rescales the CFT metric (UV side) in accordance with the normalization used in \citep{Komargodski2}, and the mode $s_p$ only contributes through total derivatives to the boundary Lagrangian. So, the QFT side is saying that once fixed the proper normalization, the corresponding coefficient of the kinetic term of the spurion should coincide with the difference of central charges. This, points towards the conclusion the modulus $\tau$ seems not to be the optimal description for the QFT spurion. In fact the PBH modulus $\tau$ looks like a warped PBH in the Fefferman-Graham gauge, see \ref{AppendixA}, so the outcome of the 2D version of the computation done in section \S 1 will change. We will show the result in the next subsection.
 
 The appropriate description of the spurion from the bulk side seems to be associated to a rigid PBH in Fefferman-Graham gauge. As we already said the modulus $\rho$ could be seen as a combination of a PBH of that kind and the mode $s_p$. So, following our line of reasoning $\rho$ seems to be the most natural bulk description of the dilaton. In fact in the 6D analysis
 to be discussed in section 4  this identification will become even more natural.
 
\subsection{Checking the PBH procedure. }

There is an equivalent way to arrive to (\ref{S2Drenormalized}). We present it here because it gives a check of the procedure
we used to compute the spurion effective action in a 4D RG flow. As was already noticed the modulus $\tau$ can be
related to a family of diffeomorphisms. To check the procedure we take as starting point the bulk on-shell action of the modulus $s_p$ without turning on $\tau$:
\begin{multline}\label{lagOnshell}
\int dtdx L^{bulk}_{2D}=\int dtdx ( \frac{g_1^2 g_2^2}{8 c_1}e^{2 s_p(t,x)}\left[y\right]^{UV}_{IR}+\left(\frac{ c_1}{2 g_1^2}\Box s_p \right)
\log
y_{UV}\\-\left(\frac{c_1(g_2^2-g_1^2)}{2 g_1^2 g_2^2}\Box s_p \right) \log y_{IR}+...+\left[L_{hom}^{\tau=0}\right]^{UV}_{IR}),
\end{multline}
and perform the UV and IR asymptotic expansions of the corresponding PBH transformation (\ref{NonWPBH}) keeping only terms
up to second order in derivatives. The
result coincides with (\ref{LagOnshell}). Notice that the PBH transformations do not affect the boundary conditions
of the matter field (\ref{vev}), provided we take the restriction (\ref{normalizability}). So all the IR constraints and
normalizability conditions we imposed before will still hold in this second approach provided they were imposed at $\tau=0$. In 

Finally, after applying the same previous procedure to the GH term and to the counterterms, namely transforming  the metric (\ref{metric}) at vanishing $\tau$-modulus,
gives (\ref{UVGH}) and (\ref{CTUV}) respectively.

\subsection{The $\rho$-branch analysis }

We can repeat the same computations done before but using the $\rho$ modulus instead of the pair $(\tau, s_p)$. The trace and off-diagonal modes $T$ and $g^{(2)}_{tx}$ can be solved from the decoupled equations $(t,t)+(x,x)$ and $(t,x)$ to be:
\begin{align}
T&=C_3(t,x)-\frac{1}{y}C_2(t,x)+\frac{c_1^2}{g_1^2 g_2^2 y \left(g_2^2\left(y+\rho\right)^2-g_1^2\rho^2\right)} \left((\partial_t \rho)^2+(\partial_x \rho)^2\right),\\
g^{(2)}_{t x}&=-\frac{C_6(t,x)}{y}+C_7(t,x)-\frac{2 c_1^2}{g_1^2 g_2^2 y \left(g_2^2(y+\rho)^2-g_1^2\rho^2\right)}\partial_t \rho \partial_x \rho.
\end{align}
In the same manner, we can then solve for all fluctuations in terms of $A^{(2)}$ by integrating  the $(t,t)-(x,x)$ and $(r,r)$ equations:
\begin{align}
\partial_y g^{(2)}&=R^{(1)}_{\partial_y g^{(2)}} A^{(2)}+R^{(2)}_{\partial_y g^{(2)}} \phi^{(2)}+\frac{1}{y^2} C_{5}, \\
\partial_y \phi ^{(2)}&=R^{(1)}_{\partial_y \phi ^{(2)}} \partial_y A^{(2)}+R^{(2)}_{\partial_y \phi ^{(2)}}\phi ^{(2)}+R^{(3)}_{\partial_y
\phi
^{(2)}}A^{(2)}+
R^{(4)}_{\partial_y \phi ^{(2)}}C_5+R^{(5)}_{\partial_y \phi ^{(2)}} \square \rho+R^{(6)}_{\partial_y \phi ^{(2)}}(\partial \rho)^2,
\end{align}
with:
\begin{align}
R^{(1)}_{\partial_y g^{(2)}}=\frac{6 g_1 g_2^3 \rho  (\rho +y)^2}{\left(g_2^2(y+\rho)^2-g_1^2\rho^2\right){}^2}, \ R^{(2)}_{\partial_y g^{(2)}}=
-\frac{2 g_1^2 \rho ^2}{(\rho +y) \left(g_2^2(y+\rho)^2-g_1^2\rho^2\right)},
\end{align}
which is also found to obey a third order linear differential equation of the form:
\begin{align}\label{ThirdOrderDiffRho}
\partial^{(3)}_y A^{(2)}+R^{(2)}_{A^{(2)}}\partial^{2}_y A^{(2)}+
R^{(1)}_{A^{(2)}}\partial_y A^{(2)}+R^{(0)}_{A^{(2)}}A^{(2)}= F_{\rho},
\end{align}
where
\begin{align}
 F_{\rho}=F^{1}(y) \Box \rho +F^{(2)}(y) (\partial \rho)^2+F^{(3)}(y)C_5(t,x).
\end{align}
The rational functions $F^{(1)}$, $F^{(2)}$ and $F^{(3)}$ are given in the second paragraph of appendix \ref{rationalFunctions}. We solve this equation by the Green's function method (see second paragraph appendix \ref{GreenF}). As for the case before we use the nine integration constants to demand IR regularity and as much normalizability as possible. In this case we are able to turn off UV sources except for one of the two corresponding to $\Delta=2$ and $\Delta=4$ CFT operators. We choose to allow  source of the $A$ scalar field, namely at the UV boundary, $y=y_{UV}$:
\begin{align}\label{sourceAA}
A^{(2)}\sim\frac{c_1^2}{3 g_1^3 g_2^3} \frac{(\partial \rho)^2-\rho \Box \rho }{ \rho^3}.
\end{align}
We compute then the full renormalized boundary action \[S_{ren}=S_{bulk}+S_{GH}+S_{CT}.\] The result up to total derivatives and without ambiguity in renormalization (as for the previous case) is:
\begin{align}\label{renAction}
S_{ren}=\int dt dx\left( \frac{c_1}{ g^2_2} \left(\partial s \right)^2+O\left(\partial^4\right)\right)
\end{align}
where $s=\log(\rho)$. Notice the coefficient in front of this kinetic term is proportional to the difference of holographic central charges among the interpolating fixed points, which in 2D can be identified with the difference of $AdS_3$  radii $\Delta L=\frac{2 c_1}{g_2^2}$. Notice that we have a freedom in normalization of $s$. We have chosen the normalization to agree with \citep{Komargodski1,Komargodski2}. Namely, the associated PBH diffeo shifts the UV/IR metric from $\eta\rightarrow e^{-2 \sigma_{PBH}} \eta$. As we mentioned the $\rho$ modulus is a combination of a PBH mode with $s_p$. So we can again check the procedure used in section \S 1 via (\ref{renAction}).


We can see the rigid $\rho$ modulus as a combination of a $PBH$ mode $y\rightarrow e^{2 \sigma_{PBH}} y$ and the $s_p=-\sigma_{PBH}$ mode. This last constraint guarantees not to turn on sources for the CFT's energy momentum tensor (nor for the hypothetical IR one). To obtain the bulk contribution we perform the PBH transformation (\ref{PBHY1})-(\ref{PBHY2}),  on the on-shell action with only $s_p$ turned on (\ref{lagOnshell}). Before performing the PBH transformation, explicit solutions in terms of $s_p$ are demanded to be IR regular and as normalizable as possible. As usual, we choose to let on the source of the dimension $\Delta=2$ CFT operator, which we can read from (\ref{sourceA}). As in previous cases. The GH and Counterterms contributions are evaluated by explicit use of the transformed metric and fields.  The GH term does not contribute to the final result for the regularized action at all. As for the CT's, they contribute with total derivatives to the final result of the effective action which, under the identification $\sigma_{PBH}\equiv s$, coincides with (\ref{renAction}).

A last comment about the relation between bulk normalizability and the identification of (\ref{renAction}) as quantum effective action for $s$: we notice that demanding normalizability of the mode $s$ amounts to impose the on-shell
condition \[\Box s=0,\] 
in both equations (\ref{sourceA}) and (\ref{sourceAA}).  This is in agreement with holographic computations of
hadron masses, where normalizability gives rise to the discreteness of the spectrum and indeed puts on-shell 
the states corresponding to the hadrons\footnote{We thank the referee for
pointing out this analogy.}.
On the other hand, the on shell supergravity  action, as already mentioned in the paragraph below (\ref{homogeneous}), is independent of $A^{(2)}$. Also,  as shown in (\ref{CTUV2}), the contributions coming from counter terms which depend on $A^{(2)}$ give  contributions that are linear in the source for the operator dual to $A$, at order $q^2$, but at  the end, these contributions reduce to a total derivatives  in (\ref{renAction}). Notice that no other sources, apart from the one corresponding to the operator dual to $A$ are turned on.
Therefore (\ref{renAction}) has no source dependence and can be interpreted as the (off-shell) effective action for the massless  mode $s$.

\section{6D Analysis}
Six dimensional supergravity coupled to one  anti-self dual tensor
multiplet, an $SU(2)$ Yang-Mills vector multiplet and one
hypermultiplet is a particular case of the general $N=1$ 6D supergravity
constructed in
\citep{Nishino}
and  admits a supersymmetric
action. The bosonic equations of motion  for the graviton $g_{MN}$,
third rank anti-symmetric tensor
${G_3}_{MNP}$, the  scalar $\theta$ and the
$SU(2)$ gauge fields $A_M^I$ are:

\begin{eqnarray}
R_{M N}-\frac{1}{2}g_{MN}R-\frac{1}{3}e^{2\theta}\big(3{G_3}_{MPQ}{G_3}_N^{\phantom{a}PQ}-\frac{1}{2}g_{MN}{G_3}_{PQR}{G_3}^{PQR}\big)\nonumber
\\-\partial_M\theta
\partial^M\theta+\frac{1}{2}g_{MN}\partial_P\theta\partial^P\theta-e^\theta
\big(2F^{IP}_MF^I_{NP}-\frac{1}{2}g_{MN}F^I_{PQ}F^{IPQ}\big)&=&0, \label{Eins}\\
e^{-1}\partial_M(eg^{MN}\partial_N\theta)-\frac{1}{2}e^\theta
F^I_{MN}F^{IMN}-\frac{1}{3}e^{2\theta}{G_3}_{MNP}{G_3}^{MNP}&=&0,
\label{thetaeq}\\
\mathcal{D}_N(ee^\theta
F^{IMN})+ee^{2\theta}G^{MNP}F^I_{NP}&=&0,\label{Feq}\\
D_M(ee^{2\theta}{G_3}^{MNP})&=&0\label{Geq}.
\end{eqnarray}
The three-form $G_3$ is the field strength of the two form $B_2$ modified
by the Chern-Simons three-form, $G_3=dB_2 + tr (F A -\frac{2}{3} A^3)$,
with the $SU(2)$ gauge field strength $F=dA+A^2$.
As a result there is the modified Bianchi identity for the 3-form:

\begin{equation}
{D}{G}_3=tr  F\wedge {F}. \label{Bianchi}
\end{equation}
We are going to consider all the fields depending on coordinates
$u$, $v$ and $r$ where $u$ and $v$ are light-cone coordinates given by $u=t+x$, $v=t-x$, and $r$ is a radial coordinate.
For the metric we take the following $SO(4)$ invariant ansatz:
\begin{eqnarray}
ds^2_6&=&e^{2f}(g_{uu}du^2 +g_{vv} dv^2+2g_{uv}du dv)+e^{-2f}\left(dr^2+r^2 d\Omega^2\right),
\label{metric}
\end{eqnarray}
where $d\Omega^2$ is the $SO(4)$ invariant metric on $S^3$:

\begin{equation}
d\Omega^2=d\phi^2 + sin^2(\phi)\left( d\psi^2
 +sin^2(\psi)d\chi^2\right),
 \end{equation}
and $f$, $g_{uu}$, $g_{uv}$, $g_{vv}$ are functions of $(u,v,r)$, from now on we will not show this dependence. As for the $SU(2)$ one-form $A$,  we take it to be non trivial only along $S^3$, preserving
a $SU(2)$ subgroup of $SO(4)$,
\begin{equation}
A=i s \sum_{k=1}^{3} \sigma^k\omega^k,
\label{gauge}
\end{equation}
where $\sigma^k$ are Pauli matrices and $\omega^k$ left-invariant one-forms  on $S^3$, and $s$ is a function of $(u,v,r)$.
For the three-form $G_3$, we take it to be non trivial only along $u,v,r$ and along $S^3$,
\begin{equation}
G_3=G^{(1)}_3 du \wedge dv \wedge  dr + G^{(2)}_3\sin^2(\phi)\sin(\psi) d\phi \wedge d\psi \wedge d\chi,
\label{threeform}
\end{equation}
where the functions $G^{(1,2)}_3$ only depend on $(u,v,r)$ . Finally we will have a non trivial  scalar  field $\theta(u,v,r)$.

\subsection{Deforming the RG flow background}
The aim of this section is to look for a solution of the above equations of motion which deforms
the RG  flow solution of \citep{Parynha},
with the appropriate boundary conditions
to be specified in due course (In order to demand IR regularity).
To be more specific, this background
is actually BPS. It preserves
half of the 8 supercharges and interpolates between two $AdS_3\times S^3$ geometries
for $r\rightarrow \infty$, the UV region,  and $ r \rightarrow 0$,  the IR region, with different
$S^3$ and $AdS_3$ radii. It describes a naively speaking, v.e.v. driven RG flow between two $(4,0)$ SCFT's
living at the corresponding $AdS$ boundaries parametrized by the coordinates $u,v$.
The solution involves an $SU(2)$ instanton centered at the origin of the  $R^4$
with coordinates $r$, $\phi$, $\psi$, $\chi$,
corresponding to $s=\rho^2/(r^2+\rho^2)$. The scale modulus $\rho$ enters  also
in the other field configurations, as will be shown shortly. Our strategy here is to
promote $\rho$ to a function of $u,v$, $\rho=\rho(u,v)$. So, the
starting point will be given by the field configurations:
\begin{eqnarray}
g_{uu}^{(0)}&=&g_{vv}^{(0)}=0, \ g_{uv}^{(0)}=-1/2,\nonumber\\
s^{(0)}&=& \rho^2/(r^2+\rho^2),\nonumber\\
f^{(0)}&=&-\frac{1}{4} log[\frac{c}{r^2}(\frac{d}{r^2}+\frac{1}{r^3}\partial_r (r^3 \partial_r log(r^2+\rho^2))],\nonumber\\
\theta^{(0)}&=&2 f^{(0)}+log(c/r^2).
\label{zeroth}
\end{eqnarray}
Notice that $s^{(0)}$ goes like $\rho^2/r^2$ in the UV.
As for the three-form, it turns out that the following expressions for $G^{(1)}_3$ and $G^{(2)}_3$ solve identically
the Bianchi identity and equations of motion:
\begin{eqnarray}\label{Gs}
G^{(1)}_3 &=&e^{4f-2\theta} \sqrt{- det g }~c/r^3,\nonumber\\
G^{(2)}_3 &=&-\left(4+d+4 s^2(-3+2 s)\right),
\label{3f}
\end{eqnarray}
where $det(g)=-g_{uu}g_{vv}+g_{UV}^2$ and $f$, $\theta$ and $s$ are functions of $(u,v,r)$.  As explained in \citep{Parynha,Duff},
the positive constants $c$ and $d$ are essentially electric and magnetic charges, respectively,
of the dyonic strings of 6D supergravity.  More precisely we have:
\begin{eqnarray}
Q_1&=&\frac{1}{8\pi^2}\int_{S^3}e^{2\theta}*G=c/4, \nonumber\\
Q_5&=&\frac{1}{8\pi^2}\int_{S^3}G=d/4+1,
\end{eqnarray}
where we see that the instanton contributes to $Q_5$ with one unit
as a consequence of the modified Bianchi identity (\ref{Bianchi}). The constants
$c$ and $d$ determine the central charges of the UV and IR
CFT's, respectively: $c_{UV}=c(4+d)$, $c_{IR}=cd$ \citep{Parynha}.

These fields solve the equations of motion only if $\rho$ is constant (apart from
$G^{(1,2)}_3$ which solve them identically). We will then deform the above background
to compensate for the back reaction due to the $u,v$ dependence of $\rho$.  In this
way one can set up a perturbative expansion in the number of $u,v$ derivatives. For the purpose
of  analyzing the equations of motion keeping track of the derivative expansion,
again it is convenient to assign a counting parameter $q$ for each $u,v$
derivative. The first non-trivial corrections to the above background
will involve two $u,v$-derivatives of $\rho(u,v)$. i.e. linear in two derivatives of $\rho(u,v)$
or quadratic in its first derivatives. From now on we will not write down the coordinate dependence of the modulus $\rho$.
Therefore we start with the following ansatz for the deformed background:
\begin{eqnarray}
f_b(u,v,r)&=&f^{(0)}(u,v,r)+q^2 f^{(2)}(u,v,r), \nonumber\\
s_b(u,v,r)&=&s^{(0)}(u,v,r)+q^2 s^{(2)}(u,v,r),\nonumber\\
\theta_b(u,v,r)&=&\theta^{(0)}(u,v,r)+q^2\theta^{(2)}(u,v,r),\nonumber\\
{g_b}_{uv}(u,v,r)&=&-1/2+q^2g_{uv}^{(2)}(u,v,r),\nonumber\\
{g_b}_{uu}(u,v,r)&=&q^2g_{uu}^{(2)}(u,v,r), \, {g_b}_{vv}(u,v,r)=q^2 g_{vv}^{(2)}(u,v,r).
\label{backq}
\end{eqnarray}
Our first task is to determine these deformations  as functions of $\rho$ and its derivatives.
The structure of the resulting, coupled differential equations for the deformations is clear:
they will be ordinary, linear second order differential equations in the radial variable $r$ with
inhomogeneous terms involving up to two  derivatives of $\rho$.
Due to the symmetry of the problem,  there is only one independent equation for the gauge field,
with free index along $S^3$, say $\phi$,
and the non trivial Einstein's equations, $E_{MN}$, arise only when $M,N$ are of type $u,v,r$ and for $M=N$
along one of the three coordinates of $S^3$, e.g. $\phi$. The traceless part of the Einstein equations $E_{uu}$ and $E_{vv}$ involve only $g_{uu}^{(2)}$
and $g_{vv}^{(2)}$ respectively and these differential equations can be solved easily. The equations $E_{uv}$, $E_{\phi \phi}$, $E_{rr}$, the gauge field equation and the
$\theta$ equation involve only
$g_{uv}^{(2)}$, $s^{(2)}(u,v,r)$, $f^{(2)}$ and $\theta^{(2)}$. Since a constant scaling of $u$ and $v$ in the zeroth order background solution is equivalent to turning on a constant
$g_{uv}^{(2)}$, the latter enters these equations only with derivatives with respect to $r$ at $q^2$ order. Therefore we can find three linear combinations
of these equations that do not involve $g_{uv}^{(2)}$. To simplify these three equations further, it turns out  that
an algebraic constraint among the fields $f$, $\theta$ and $s$,
dictated by consistency of the $S^3$ dimensional reduction of the 6D theory down
to 3D, gives a hint about a convenient way to decouple the differential equations by redefining the field $\theta$
in the following way.
\begin{equation}
e^{\theta}=\frac{r^2 e^{-2 f}e^{\varphi}}{(4+d-s^2)}.
\label{constraint}
\end{equation}
Note that for the reduction ansatz, $\varphi=0$. In general the new field $\varphi$ will also have an expansion in $q$
of the form:
\begin{equation}
\varphi(u,v,r)=\varphi^{(0)}(u,v,r)+q^2 \varphi^{(2)}(u,v,r).
\end{equation}
For the zeroth order solution defined above one can see that $\varphi^{(0)}=0$. The reduction ansatz indicates that
at order $q^2$ one can find a combination of the linear second order differential equations which gives a decoupled
homogeneous second order equation for
$\varphi^{(2)}$. This equation can be solved for $\varphi^{(2)}$, which involves two integration constants denoted by $a_1$ and $a_2$ (that are functions of $u$ and $v$)
\begin{eqnarray}
 \varphi^{(2)}_h &=& a_1(u, v)\frac{48 r^6(r^2+\rho^2)^2   \log(\frac{r^2 + \rho^2}{r^2}) - 48 r^6 \rho^2 - 24 r^4 \rho^4 + (12 + d) r^2 \rho^6 + d \rho^8 }{r^4 \rho^2 ((4 + d) r^4 + 2 (4 + d) r^2 \rho^2 +
d \rho^4)} \nonumber\\
~~~ &+&a_2(u, v)\frac{4r^2 (r^2 + \rho^2)}{\rho^2 ((4 + d) r^4 + 2 (4 + d) r^2 \rho^2 + d\rho^4)},
\end{eqnarray}
and after substituting this  solution, we get two second order differential equations for $s^{(2)}$ and $f^{(2)}$. In general one can eliminate $f^{(2)}$ from these two equations and obtain a fourth order
differential equation for $s^{(2)}$. However, it turns out that in these two equations $f^{(2)}/r^2$ appears only through $r$-derivatives \footnote{This can be understood by observing that one can add  a constant to
the solutions  for $e^{\theta-2 f}$ and $e^{-\theta-2 f}$ in equations (3.28) and (3.26). At the infinitesimal level this is equivalent to turning on a constant  $f^{(2)}/r^2$.}
and this results in a third order decoupled differential equation for $s^{(2)}$
\begin{equation}
 A_3(r) \partial_r^3 s^{(2)}+ A_2(r)\partial_r^2 s^{(2)}+ A_1(r)\partial_r s^{(2)}+A_0(r) s^{(2)} = B(r),
\end{equation}
where
\begin{eqnarray}
A_3(r) &=& r^3 (r^2 + \rho^2)^6 ((4 + d) r^4 +
   2 (4 + d) r^2 \rho^2 + d \rho^4)^2, \nonumber\\
A_2(r) &=&  r^2 (r^2 + \rho^2)^5 (11 (4 + d)^2 r^{10} +
   51 (4 + d)^2 r^8 \rho^2 +
   2 (4 + d) (128 + 47 d) r^6 \rho^4 \nonumber\\
   &~&~+
   2 (4 + d) (24 + 43 d) r^4 \rho^6  +
   d (80 + 39 d) r^2 \rho^8 + 7 d^2 \rho^{10}),\nonumber\\
 A_1(r)&=& r (r^2 + \rho^2)^4 (21 (4 + d)^2 r^{12} +
   130 (4 + d)^2 r^{10} \rho^2 + (4 + d) (948 + 311 d) r^8 \rho^4 \nonumber\\
   &~&~+
   4 (4 + d) (100 + 91 d) r^6 \rho^6 +(-192 + 456 d + 211 d^2) r^4 \rho^8 +
   10 d (-8 + 5 d) r^2 \rho^{10} + d^2 \rho^{12}),\nonumber\\
 A_0(r)&=&\ 16 \rho^2 (r^2 + \rho^2)^3 (4 (4 + d)^2 r^{12} + (4 + d) (72 + 19 d) r^{10} \rho^2 +(4 + d) (72 + 35 d) r^8 \rho^4 \nonumber\\
 &~&~+
   2 (16 + 54 d + 15 d^2) r^6 \rho^6 +
   2 d (6 + 5 d) r^4 \rho^8 - d^2 r^2 \rho^{10} -
   d^2 \rho^{12}), \nonumber\\
 B(r)&=&  16 c \rho (r^2 + \rho^2)^2 ((4 + d) r^4 +
   2 (4 + d) r^2 \rho^2 + d \rho^4)^3 \partial_u \partial_v \rho \nonumber\\
   &~&~+ 16 c (r^4 + 2 r^2 \rho^2 -
   3 \rho^4) ((4 + d) r^4 + 2 (4 + d) r^2 \rho^2 +
    d \rho^4)^3 \partial_u \rho \partial_v \rho.
\end{eqnarray}

The three independent solutions of the homogeneous part of the above equation are
\begin{eqnarray}
s^{(2)}_h&=&a_3(u,v)\frac{3 (4 + d) r^8 +
  24 (4 + d) r^6 \log(r/\rho) \rho^2 -
  6 (10 + 3 d) r^4 \rho^4 - 6 (2 + d) r^2 \rho^6 -
   d \rho^8}{12 r^4 (r^2 + \rho^2)^2}+\nonumber\\
&~&a_4(u,v)\frac{ \rho^2(24 r^6 \log(1 + \rho^2/r^2) -
 24 r^4 \rho^2 + 3 (8 + d) r^2 \rho^4 +
 2 d \rho^6)}{144 r^4 (r^2 + \rho^2)^2}
+a_5(u,v)\frac{r^2 \rho^2}{(r^2 + \rho^2)^2}.\nonumber \\
\label{sh}
 \end{eqnarray}
Using the most general solution of the homogeneous equation one can construct the Green's function for the third order differential equation  and obtain a particular solution of the full inhomogeneous equation
 \begin{eqnarray}
s^{(2)}_p&=&~\frac{c (3 (4 + d) r^6 - 6 (4 + d) r^4 \rho^2 -
    2 (30 + 7 d) r^2 \rho^4 - 5 d \rho^6)}{3 r^4 (r^2 + \rho^2)^3}\partial_u \rho\partial_v \rho \nonumber\\&~&~~~~~+\frac{c\rho  (3 (4 + d) r^4 +
   3 (4 + d) r^2 \rho^2 + d \rho^4)}{3 r^4 (r^2 + \rho^2)^2}\partial_u \partial_v \rho.
 \label{sp}
 \end{eqnarray}
 
Substituting the general solution for $s^{(2)}$ in the remaining equations one gets first order linear differential equations for $f^{(2)}$ and $g_{uv}^{(2)}$
which can be solved easily resulting in two more integration constants. Moreover, $E_{uu}$ and $E_{vv}$ give two decoupled second order differential equations for the traceless part of the
metric $g_{uu}^{(2)}$ and $g_{uu}^{(2)}$ that can also be readily solved giving another four integration constants. In all there are eleven integration constants as compared to nine
integration constants in the 3D case discussed in the previous sections. This is to be expected since the $S^3$ reduction ansatz from 6D to 3D sets $\varphi=0$. Finally $E_{ru}$ and $E_{rv}$ at
order $q^3$ give first order partial differential equations in $u$ and $v$ variables on the integration constants. The full homogeneous solution and a particular solution for the inhomogeneous equations are given in Appendix \ref{AppD}.

Now we turn to the analysis of the IR and UV behaviour of the general solutions. The general solution for $s^{(2)}$ is a sum of the particular solution (\ref{sp}) and the homogeneous solution (\ref{sh}).
Near $r=0$ this solution has divergent $1/r^4$ and $1/r^2$ terms that can be set to zero by choosing:
\begin{equation}
a_3(u, v)=\frac{4 c \partial_u \partial_v \log \rho}{3 \rho^2},~~~~
a_4(u,v)=\frac{16 c}{ \rho^4}(7\partial_u\rho\partial_v\rho-\rho\partial_u \partial_v \rho).
\end{equation}
Similarly analyzing the general solution for $\varphi^{(1)}$ one finds that it has also IR divergent $1/r^4$ and $1/r^2$ terms that can be set to zero by setting $a_1(u,v)=0$. With these choices we have checked that
Ricci scalar and Ricci square curvature invariants are non-singular at $r=0$.

Finally, the Einstein equations $E_{ur}$ and $E_{vr}$ give certain partial differential equations with respect to $v$ and $u$ on the integration constants $b_1$ and $c_1$ respectively and these are
solved by:
\begin{equation}
 b_1= \frac{4 c \left(-2(\partial_u\rho)^2+ \rho \partial_u^2 \rho \right)}{\rho^2}, ~~~~~c_1= \frac{4 c\left(-2(\partial_v\rho)^2+ \rho \partial_v^2 \rho\right)}{\rho^2}.
\end{equation}
With these conditions even the metric functions $g_{uu}$, $g_{vv}$ and $g_{uv}$ have no power like singularities in $r$ near $r \rightarrow 0$. Thus we have a smooth solution near IR
up to $q^2$ order.

In the UV region, $r\rightarrow \infty$, the source terms behave as $O(r^2)$ for $\varphi$ and $f$, and $O(1)$ for the metric $g_{uv}$, $g_{uu}$ and $g_{vv}$. By making an asymptotic expansion
of the homogeneous solutions one can see that $a_2$, $a_4$, $a_7$, $b_2$ and $c_2$ control these source terms. Since in our background we do not want to turn on any sources, we set these
integration constants to zero.



Finally, the UV behaviour of the gauge field $s^{(2)}$ is:
\begin{equation}
 \frac{c(4+d) \partial_u\partial_v \log \rho}{3 \rho^2} (1 - \frac{2 \rho^2}{r^2} (4 \log(\frac{\rho}{r}+1)) +\frac{\rho^2}{r^2} a_5.
\end{equation}

It turns out though that IR regularity forces us  to allow a source term for the $s_b(u,v,r)$ field:
this is a term of order $r^0$ for $r\rightarrow\infty$, of order $q^2$:
\begin{equation}
s^{(2)}\rightarrow \frac{c(4+d)(\rho\partial_u\partial_v\rho-\partial_u\rho\partial_v\rho)}
{3\rho^2}+{\mathcal{O}}(1/r),
\label{source}
\end{equation}
 as $ r\rightarrow\infty$.
Notice that here, like in the 3D case,  discussed at the end of section \S 3, the source term for the operator dual to $s$ 
 is proportional to the EoM for the massless scalar $\log \rho$, and therefore vanishes on-shell.

\subsection{Finding linearized fluctuations around the deformed background}
Having determined the background corrected by the leading terms involving  two space-time
derivatives of the modulus $\rho$, we could compute the regularized on shell action,
as was done in the 3D case. We find it more convenient to compute directly one-point functions
of dual operators (especially of the stress energy tensor). To this end we need
to switch on corresponding sources and therefore to solve the linearized equations
of motion of the various fields on the deformed background. This is done
again in a derivative expansion starting with the following ansatz for the fields fluctuations:
\begin{gather}
\delta s=\delta^{(0)} s+q^2 \delta^{(2)} s, \ \delta g_{uu}=\delta^{(0)} g_{uu}+q^2 \delta^{(2)}g_{uu}, \delta g_{vv}=\delta^{(0)} g_{vv}+q^2 \delta^{(2)}g_{vv}, \\ \delta g_{uv}=\delta^{(0)} g_{uv}+q^2 \delta^{(1)}g_{uv}, \
\delta f =\delta^{(0)}f +q^2 \delta^{(2)}f, \ \delta \theta=\delta^{(0)}\theta +q^2 \delta^{(2)}\theta,
\end{gather}
where $\delta^{(0)}$ stands for the zeroth order in space-time derivatives, and $\delta^{(2)}$ stands for fluctuations coming at second order in space time derivatives
and this is why is weighted by $q^2$.  The  general solution for $\delta^{(0)}$ is the homogeneous solution given in Appendix \ref{AppD}.  We fix the integration constants so that
\begin{gather} \label{dzeroth}
\delta^{(0)}g_{uu}=h_{uu}   , \ \delta^{(0)} g_{vv}=h_{vv},  \ \delta^{(0)} g_{uv}=h_{uv}, \\
\delta^{(0)}f=\frac{2 \rho ^4 r^2 }{\left(\rho ^2+r^2\right) \left(d \rho ^4+\left(4+d\right) r^4+2 \left(4+d\right) \rho ^2 r^2\right)}a_{5}(u,v), \\
\delta^{(0)}\theta = \frac{4  \rho ^4 r^2}{\left(\rho ^2+r^2\right) \left(d \rho ^4+\left(4+d\right) r^4+2 \left(4+d\right) \rho ^2 r^2\right)}a_5(u,v),\\
\delta^{(0)} s=\frac{r^2 \rho^2 a_5(u,v)}{(r^2+\rho^2)^2},
\end{gather}
where $h_{uu}$, $h_{vv}$ and  $h_{uv}$ are the integration constants $b_2(u,v)$, $c_2(u,v)$ and $a_7(u,v)$  respectively. Consequently they are the sources for the boundary stress energy tensor components $T_{uu}$, $T_{vv}$, and $T_{uv}$.
These $h$'s are small fluctuations around the flat boundary metric, $g^{(0)}=\eta+h$, and the corresponding linearized curvature is
\begin{equation}
R^{(2)}(g^{(0)})=-2(\partial^2_vh_{uu}-2 \partial_u \partial_v h_{uv}+\partial^2_u h_{vv}).
\label{2dcurv}
\end{equation}
We have also kept the integration constant $a_{5}$ for reasons that will become apparent later on.

The next step is to solve the equations of motion at order $q^2$ for the $\delta^{(2)}$ fields. The equations for  $\delta^{(2)}$ fields contain also inhomogeneous terms
that involve $\delta^{(0)}$ fields and their derivatives, up to second order with respect to $u$ and $v$.
The procedure is the same
as the one employed in  solving for the corrected background. As the differential equations are inhomogeneous, the general solution
will be the sum of the homogeneous solution and a particular solution
of the inhomogeneous one, which can be obtained using Green's functions once we have the homogeneous solutions.
The integration constants in the homogeneous part of solution can be partially fixed by requiring IR smoothness and absence of sources for
$ \delta^{(2)}\theta$ and $ \delta^{(2)} f$. Moreover some sources can be reabsorbed in the already existing
sources at zeroth order. Finally, the mixed $u,v$ and $r$ Einstein's equations result in differential constraints among the integration
constants.


Concerning the IR behaviour, the metric components go,  for $r\rightarrow 0$, as:
\begin{eqnarray}
\delta^{(2)}g_{uv}&\sim& -\frac{c d}{4}R^{(2)}/r^2,\nonumber\\
\partial_v\delta^{(2)}g_{uu}&\sim&-\frac{c d}{4}\partial_u R^{(2)}/r^2,\nonumber\\
\partial_u\delta^{(2)}g_{vv}&\sim&-\frac{c d}{4}\partial_v R^{(2)}/r^2.\nonumber\\
\end{eqnarray}
The apparent $1/r^2$ singularity is presumably a coordinate singularity:
we have verified that both the 6D Ricci scalar and Ricci squared are finite both at the IR and UV.
The other fields are manifestly regular at the IR.
We have seen that the there is a physical fluctuation for the operator ${\cal O}_s$   
proportional to $\rho^2$ at  order $q^0$
and that at order $q^2$ there is a source, $J_s$, which couples to it, proportional to $\Box \log(\rho)/\rho^2$.
Therefore we expect that, at order $q^2$,  the corresponding term ${\cal O}_s J_s$ in the boundary action
will not give any contribution being a total derivative. So, this type of term will not contribute to the dilaton
$\rho$ effective action if we were to compute it, as it was done in the 3D case, by evaluating the regularized
bulk action on the background together with boundary GH and counter-terms.
We close this subsection by  writing down the full source term $J_s$  for the operator ${\cal O}_s$  dual
to the bulk field $s$, i.e. the sum of the source in the background  $s_b$ plus the one in the fluctuation $\delta s$:

\begin{multline}
J_s=\frac{c(4+d)}{12}\left(\frac{\Box_{g^{(0)}} \log(\rho)-\frac{1}{2}R^{(2)}[g^{(0)}] }{\rho^2} \right)\\
+ \frac{c(4+d)}{12}\frac{\Box \log(\rho)}{\rho^2}a_{5}+\frac{c(4+d)}{24}\frac{1}{\rho^2}\Box a_{5}.
\label{ssource}
\end{multline}
Next, we go to compute the contribution of the term
$\int\sqrt{g^{(0)}} J_s {\cal{O}}_s$
to the 2D boundary action. While $ J_s$ is the coefficient of $r^0$ in the UV expansion of
$s$, $< {\cal{O}}_s>$ is proportional to the coefficient of $1/r^2$. We will determine this
proportionality constant in the following by studying the dependence of the regularized bulk action on $a_{5}$. Note that
$ J_s$ is already of order $q^2$, therefore we need only $q^0$ term in the coefficient of $1/r^2$ in $s$, which can be seen from (\ref{zeroth}) and (\ref{dzeroth}) to be\footnote{Of course, the same remarks about the CFT interpretation of the asymptotic
data of bulk fields made in sub-section \S 3.1, implying spontaneous symmetry breaking of conformal invariance, apply here.}
\begin{equation}\label{vev0}
 < {\cal{O}}>_s \sim \rho^2(1+a_{5}).
\end{equation}
Using the fact that $\sqrt{g^{(0)}}$ at order $q^0$ is $1/2(1-2 h_{uv})$, it can be shown that
$\sqrt{g^{(0)}} J_s < {\cal{O}}_s>$ up to the order we are working at, is a total derivative and therefore the corresponding
integral vanishes.

\subsection{Boundary Action}

Here, we will determine  the boundary action in presence of sources
for the dual stress energy tensor $T_{\mu\nu}$ , which will allow to compute
its one-point functions.
We will expand the bulk action around the determined background
 to linear order in the fluctuation fields, at order $q^2$.
First of all, we need to point out a subtlety concerning the bulk action.
Recall that the bosonic  equations of motion of $(1,0)$ 6D supergravity, (\ref{Geq}), can be derived
from the following action:

\begin{align}\label{6DAction}
S^{bulk}_{6D}=\int d^6 x\sqrt{-g_{6D}}\left(-\frac{1}{4} R+\frac{1}{4}e^{\theta}F^2-\frac{1}{4}e^{2 \theta}(G_3)^2-\frac{1}{4} (\partial \theta)^2  \right),
\end{align}
where the equations of motion are obtained by varying with respect to all the
fields, including the two form $B_{MN}$. The 6D equations of motion have been
shown in
\citep{Parynha}
to reduce consistently to the 3D equations discussed earlier. In particular
the 3D flow solution discussed before has a 6D uplift. For convenience, we give the map of the  6D fields and parameters in terms of  3D ones used in the previous sections:
\begin{eqnarray}
&~&r^6 e^{-8 f} dr^2 \rightarrow dr^2,~~~ r^3 e^{-2f}\rightarrow e^f,~~~ s\rightarrow 2 A,~~~e^{4\theta}\rightarrow \frac{g_1^6 e^{ 2\phi}}{256 g_2^8\left(1-A^2\right)^{3}},\nonumber \\
&~& \varphi \rightarrow  0,~~~~4+d\rightarrow \frac{4 g_2^2}{g_1^2},~~~~ \ c \rightarrow\frac{c_1}{2g_2^2}.
\end{eqnarray}

In the 6D action (\ref{6DAction}) above,  $(G_3)^2$ equals $(G_3^{(1)})^2+(G_3^{(2)})^ 2$. However the  3D gauged supergravity action is not the reduction of $S^{bulk}_{6D}$. The difference lies in the fact that
in reducing to 3D, one eliminates $G_3$ by using its 6D solution in terms of the remaining fields. The 3D action is constructed by demanding that its variation gives the correct  equations for the remaining fields.
From the
explicit solutions for $G_3^{(1)}$ and $G_3^{(2)}$ in (\ref{Gs}), one can easily prove that the modified action $\tilde{S}^{bulk}_{6D}$, obtained by replacing $(G_3)^2\rightarrow (G_3^{(1)})^2-(G_3^{(2)})^2$ in $S^{bulk}_{6D}$,
reproduces the correct equations of motion for all the remaining fields. From the AdS/CFT point of view, it seems reasonable to use $\tilde{S}^{bulk}_{6D}$, since the two-form potential in 3D is not a propagating
degree of freedom and does not couple to boundary operators.  We should point out that the boundary action that we will compute in the following is not the same for $S^{bulk}_{6D}$ and $\tilde{S}^{bulk}_{6D}$. Only the latter
reproduces the results of the 3D analysis. The flow solution studied in this paper can be described in the 3D gauged supergravity, however there are many solutions describing flows in 2D or 4D CFTs that
cannot be described  in 3D or 5D gauged supergravities. Instead one has to directly work in higher dimensions. In such cases, we think, that the bulk action that should be used in the holographic computations, is the one that
reproduces the correct equations for the fields that couple to the boundary operators, after having eliminated 2-form and 4-form fields respectively.

As promised at the beginning of this subsection our goal will be to evaluate $S^{bulk}_{6D}$, with the
modification just mentioned, on the field configurations which are sums of the
background fields plus the $\delta$ fields, at first order in the latter and to order $q^2$. Since
the background solves the equations of motion, the result will be a total derivative and there will
be possible contributions from the UV and IR boundaries, i.e.  $r\rightarrow \infty$
and $r\rightarrow 0$, respectively. It is simpler to  give the sum, $S_1$,  of the boundary term coming
from the bulk action and  the Gibbons-Hawking term, which in our case is $\int du dv \frac{\partial_r\left(e^{2f} det(g)\right)}{\sqrt{(-det g)}}$:
\begin{eqnarray}
S_1&=&\int \frac{du dv} {\sqrt{-det g}} [-r^3(4{g_b}^2_{uv}\partial_r f_b-{g_b}_{vv}\partial_r{g_b}_{uu}+2{g_b}_{uv}\partial_r{g_b}_{uv}\nonumber\\
&-&{g_b}_{uu}(4{g_b}_{vv}\partial_rf_b+\partial_r{g_b}_{vv}))\delta f/2\nonumber\\
&-&r^2({g_b}_{uv}(-6+4r \partial_rf_b)-r\partial_r{g_b}_{uv})\delta g_{uv}/4\nonumber\\
&+&r^2({g_b}_{vv}(-6+4r\partial_rf_b)-r\partial_r{g_b}_{vv})\delta g_{uu}/8\nonumber\\
&+&r^2({g_b}_{uu}(-6+4r\partial_rf_b)-r\partial_r{g_b}_{uu})\delta g_{vv}/8\nonumber\\
&-&r^3(-det g )\partial_r \theta_b\delta\theta\nonumber\\
&-&6e^{2f_b+\theta_b}{(-det g)}\delta s].
\label{S1}
\end{eqnarray}

By looking at the solutions for the various fields one can see that this expression
has a quadratic divergence for $r\rightarrow \infty$ at order $q^0$, which can be renormalized
by subtracting a counterterm proportional to the boundary cosmological constant:
\begin{equation}
S_{CT}= \frac{1}{2(c(4+d))^{1/4}}\int dudv e^f \sqrt{-det g}.
\label{CT}
\end{equation}
The final term $S_f=S_1-S_{CT}$, at order $q^2$,  for $r\rightarrow \infty$ is obtained
using the explicit solutions:

\begin{eqnarray}
S_f&=&\int du dv \frac{c}{8\rho^2}(2 h_{uu}(9(\partial_v\rho)^2-\rho\partial^2_v\rho)\nonumber\\&+&
2\partial_u\rho(8 a_5\partial_v\rho-16 h_{uv}\partial_v\rho+9 h_{vv}\partial_u\rho\nonumber\\&+&
2\rho(7\partial_u\rho\partial_u h_{vv}-8 a_5\partial_u\partial_v\rho+16 h_{uv}\partial_u \partial_v\rho\nonumber\\
&+& h_{vv}\partial^2_u\rho)+7\rho^2(\partial_{vv} h_{uu} +\partial_u\partial_v a_5-2\partial_u\partial_v h_{uv}+\partial^2_u h_{vv}))).
\label{final}
\end{eqnarray}
For $r\rightarrow 0$ one can readily verify that there is no finite contribution left over.
Before coming to the computation of $<T_{uu}>$,  $<T_{vv}>$ and $<T_{uv}>$, let us analyze more
precisely ${\cal O}_s$ . This can be obtained by comparing $J_s$ from (\ref{ssource}), after setting to zero
the sources of $T_{\mu\nu}$,
with the corresponding term in $S_f$, which gives $\int \sqrt{g^{(0)}} <{\cal O}_s>J_s$.
Setting the sources of $T_{\mu\nu}$ to zero, i.e. keeping only  $ a_5 $, $S_f$  is $2c (\partial_u\rho\partial_v\rho-\rho\partial_u\partial_v\rho)/\rho^2 a_5 $ which by the
holographic map  is equal to $\int \sqrt{g^{(0)}} <{\cal O}_s >J_s$. Using the expression for  $J_s$ given in (\ref{ssource}) one finds:
\begin{equation}
<{\cal O}_s>^{(0)}=\frac{6\rho^2}{4+d}.
\label{vev1}
\end{equation}
Notice that using the fact that $<{\cal O}_s>$ is proportional to $\rho^2$ (\ref{vev0}), the term proportional to $\Box a_5$ in $J_s$ is  a total derivative.  The above equation actually gives the
proportionality constant in (\ref{vev0}) so that including the first order fluctuation:
\begin{equation}
<{\cal O}_s>=\frac{6\rho^2}{4+d}(1+a_5).
\label{vev2}
\end{equation}

\subsection{One-point function of $T_{\mu\nu}$}
The one-point functions of the stress energy tensor, $<T_{uu}>$, $<T_{vv}>$
and $<T_{UV}>$, are determined
as the coefficients of $h_{vv}$, $h_{uu}$ and $h_{uv}$, respectively, in $S_f$. After performing a partial integration one
obtains the result :
\begin{eqnarray}
<T_{uu}>&=&\frac{-2 c(2(\partial_u\rho)^2+\rho \partial^2_u\rho)}{\rho^2},\nonumber\\
<T_{vv}>&=&\frac{-2 c(2(\partial_v\rho)^2+\rho \partial^2_v\rho)}{\rho^2},\nonumber\\
<T_{uv}>&=&\frac{2 c(-\partial_u\rho\partial_v\rho+\rho\partial_u\partial_v\rho)}{\rho^2}.
\label{tensor}
\end{eqnarray}
This stress energy tensor can be derived from an effective action for the field $\rho$:

\begin{equation}
S_\rho=2c\int dudv\sqrt{- g^{(0)}}[( \partial \log(\rho))^2-R^{(2)}(g^{(0)}) \log(\rho)].
\label{dilaton}
\end{equation}
Note that the coefficient that appears in $S_\rho$ is $c$ which is proportional to $c_{UV}-c_{IR}$.
Under the Weyl transformation
\begin{equation}
g^{(0)}\rightarrow  e^{2 \sigma} g^{(0)}, ~~~~~\rho\rightarrow  e^{-\sigma} \rho,~~~~~S_\rho \rightarrow S_\rho + 2c\int dudv\sqrt{- g^{(0)}}R^{(2)} g^{(0)},
\end{equation}
and therefore $S_\rho$ precisely produces the anomalous term. Finally note that $J_s$  in (\ref{ssource}) transforms, up to the linearized fluctuation that we have computed here,
covariantly as  $J_s\rightarrow e^{2 \sigma} J_s$ under the Weyl transformation.

Finally, using (\ref{tensor}), (\ref{ssource}) and (\ref{vev2}), we find that the conservation of stress tensor is modified by the source terms as:
\begin{equation}
\partial^i< T_{ij}> = J_s \partial_j <{\cal O}_s>,
\end{equation}
which is the Ward identity for diffeomorphisms in the CFT in the presence of a source term $\int j_s{\cal O}_s$.

Now we would like to interpret (\ref{dilaton}) from the dual $(4,0)$ SCFT point of view.  It is useful to
recall some facts from the better understood type IIB $(4,4)$ SCFT describing bound states of $Q_1$ D1-branes and
$Q_5$ D5 branes \citep{SW, Aharony}. If one wants to study the
separation of, say, one D1 or D5 brane from the rest, one has to study the effective action for
the scalars in the vector multiplets, $\vec{V}$, in the relevant branch of the 2D (4,4) gauge theory, which is the Higgs branch, where (semiclassically)
the hypermultiplet scalars $H$ acquire v.e.v., whereas  for the vector multiplet scalars,
which carry dimensiion 1,  $<V>=0$. One can obtain an effective action for $V$ either by a probe supergravity approach \citep{SW} or by a field theory argument \citep{dps,witten,Aharony}, i.e. by integrating out the hypermultiplets and observing that in the 2D field theory
there is a coupling schematically of the form ${\vec{V}}^2 H^2$. This can be shown to produce for $\log |\vec{V}|$ a lagrangian of the form (\ref{dilaton}) with the correct background charge to produce a conformal anomaly which  matches the full conformal anomaly, to leading order in the  limit of large charges.

In our case, where we have a D1-D5 system in presence of D9 branes in type I theory,
the role of the vector multiplet scalars is played by the field $\rho$,  the instanton scale in the background geometry.
The "separation" of one D-brane corresponds geometrically to the limit $\rho\rightarrow\infty$, where the gauge 5-brane
decouples, making a reduction in the central charge from an amount proportional to $Q_1 Q_5$ in the UV to $Q_1 ({Q_5}-1)$ in the IR, where, as shown earlier, $Q_1=c/4$ and $Q_5=d/4+1$. Therefore the variation of the central charge,  $\delta c$,  is proportional to $Q_1$.
On the other hand, from the D-brane effective field theory point of view the instanton scale corresponds to a gauge invariant combination of the D5-D9  scalars, $h$, with $h^2\sim \rho^2$. The $h$'s  are in the bifundamental of $Sp(1)\times SO(3)$, $Sp(1)$ being the gauge group on the 
D5-brane and $SO(3)$ that on the D9- branes.
The $h$'s couple to  D1-D5 scalars $H$  which are in the bifundamental of $SO(Q_1)\times Sp(1)$ and belong to (4,4) hypermultiplets. In the Higgs branch, which gives the relevant dual
CFT, again $H$'s can have v.e.v. semiclassically, while $<h>=0$. In
the 2D effective action there is a coupling of the form $H^2 h^2$ and upon 1-loop integration 
of $H$'s one gets a term $(\partial h)^2/h^2$\citep{dps}, with coefficient proportional to $Q_1$.
The presence of the background charge term can be justified along the lines of \citep{SW, Aharony}.

\section{Conclusions and Open Problems}

This article consists of two parts. In the first part, section \S 2, we have shown how Weyl anomaly matching 
and the correspondig Wess-Zumino action for the "spurion" is
reproduced holographically, from kinematical arguments on the bulk gravity side: there, its universality comes from the fact that only the leading  boundary behaviour of bulk fields enters the discussion. The PBH diffeomorphisms affect the boundary data and consequently the gravity action depends on them, in 
particular on the field $\tau$. The regulated effective action  is completely fixed by the kinematical procedure detailed in section \S 1. For a specific representative in the family of diffeomorphisms the Wess Zumino term takes the minimal form reported in literature. 
In appendix \ref{AnomalyM} we present a different way to approach the same result (We do it for an arbitrary background metric).

We  then moved on in sections \S 3 and \S 4  to analyze  an explicit  3D holographic RG flow solution, 
which has a "normalizable" behaviour  in the UV. In section \S 3 we studied the problem in the context of 3D gauged 
supergravity.
 We started by identifiying the possible moduli of the background geometry: out of the zero modes $(\tau, s_p, \rho)$, there come out two independent normalizable combinations. We promoted these integration constants to 
functions of the boundary coordinates $(t,x)$ and solve the EoM up to second order in a derivative expansion.
In a first approach we used a combination of $(\tau, s_p)$ dictated by normalizability, in a second approach we 
used $\rho$. In both cases we find a boundary action for a free scalar field with the expected normalization. As argued in section \S 3, agreement with QFT arguments in \citep{Komargodski2} points towards $\rho$ as the right description for the would-be-dilaton scalar field.  For possible extensions to higher dimensional computations, could be helpful to keep on mind that this mode $\rho$ can be seen as the normalizable combination of a rigid PBH in Fefferman-Graham gauge and the mode $s_p$.

Then we moved in section \S 4 to elucidate the QFT interpretation of this 
normalizable mode by lifting the 3D theory to the 6D one:
we promoted the modulus $\rho$, the $SU(2)$ instanton scale, to a boundary field, $\rho(u,v)$,
and  solved the EoM in a derivative expansion both for the background geometry and the linearized
fluctuations around it, up to second order. This allowed us to compute $<T_{\mu\nu}>$ and
determine  the boundary action for $\log \rho$: this is the action of a free scalar with background charge
and its conformal anomaly is $c_{UV}-c_{IR}$, therefore matching the full $c$. 
We identified $\tau=log\rho$ with a D5-D9 mode in the $(4,0)$ effective field theory of the
D1-D5 system in the presence of D9 branes in type I theory.

Finally, as an open problem, it would be interesting to apply the procedure 
followed in sections \S 3 and \S 4 to a v.e.v. driven RG flow in a 5D example, 
where we would give spacetime dependence to the moduli associated, say,  to the Coulomb branch of 
a 4D gauge theory: in this case no subtleties related to spontaneous 
symmetry breaking arise and we should be able to obtain a genuine dilaton effective action.



\label{sec:Conclusion}
\acknowledgments

We would like to thank Adam Schwimmer for his comments and
Fernando Quevedo for reading the manuscript. KSN acknowledges partial support from the European Commission under contract PITN-GA-2009-237920. ACB acknowledges The Abdus Salam ICTP for financial support and hospitality.  

\clearpage

\appendix

\section{PBH diffeomorphisms}

\subsection{Conventions}

We use the mostly positive convention for the metric, namely signature $(-,+,+,+)$ in 4D and $(-,+)$ in 2D.
The Riemann tensor we define as: \[{R_{\mu \nu \alpha}}^{\beta}=2 \partial_{[\mu} \Gamma^{\beta}_{\nu] \alpha}+2 \Gamma_{[\mu \lambda}^{\beta} \Gamma_{\nu] \alpha}^{\lambda}, \]
with the Christoffel symbols:
\[
\Gamma^{\beta}_{\nu \alpha}=\frac{1}{2}g^{\beta \eta}\left( \partial_\nu g_{\eta \alpha}+ \partial_\alpha g_{\eta \nu}-\partial_\eta g_{\nu \alpha}\right).
\]
The 4D Euler density and Weyl tensors are defined as:
\begin{align}
E_4=R^2_{\mu \nu \rho \sigma}-4 R^2_{\mu \nu}+R^2, \ C = R^2_{\mu \nu \rho \sigma}-2 R^2_{\mu \nu}+ \frac{1}{3}R^2.
\end{align}


\subsection{Non Static domain wall ansatz} \label{AppendixA}
Be the domain wall form for the metric:
\begin{align}
ds^2=dr^2+e^{2 f(r,x)}g_{\mu \nu}(x,r)dx^{\mu}dx^{\nu}.
\end{align}
The PBH diffeomorphism until second order in derivatives of $\tau$, can be written by symmetry arguments are:
\begin{align}\label{NonWPBH}
 x^{\mu}\rightarrow &x^{\mu}-a^{(1)}[r+\tau,x] \partial^{\mu} \tau+O\left(\partial^3\right)\\
 r \rightarrow & r+\tau+b^{(3)}[r+\tau,x] (\partial \tau)^2+O\left(\partial^4 \tau\right),
\end{align}
where index contractions and raising up of covariant indices are made by using the metric $g^{\mu \nu}=g^{\mu \nu}(r,x^\mu)$. The gauge preserving conditions on the form factors are
\begin{align}
\partial_z a^{(1)}[z,x]=e^{-2f}, \ \partial_z b^{(3)}[z,x]= \frac{e^{2 f}}{2}(\partial_z a^{(1)})^2,
\end{align}
where $z=r+\tau$. Notice that if we go to the Fefferman-Graham gauge this mode will look like a "warped" diffeomorphism. Namely, the induced
y-transformation at zeroth order in derivatives of $\tau$ will look like: \[y\rightarrow y e^{ h(y) \tau },\] with $h$ some function of $y$ interpolating between constant values. This is the technical cause behind the fact the coefficient in the kinetic term (\ref{S2Drenormalized}) does not coincides with the difference of holographic central charges. Namely, if we choose the right normalization in the UV $h(\infty)=1$, thence $h(0)\neq1$, and so the IR kinetic contribution is not properly normalized to the IR central charge.

\subsection{Non Static Fefferman Graham gauge}\label{AppendixA2}
Let us suppose we are in the Fefferman-Graham gauge, namely:
\begin{align}
ds^2=g_{yy}(y) \frac{dy^2}{y^2}+ y \left( g_{\mu \nu}dx^\mu dx^\mu \right),
\end{align}
where $g_{yy}$ and $g_{\mu \nu}$ go like constant and a space time function times $\eta_{\mu \nu} $ in both UV and IR limits, respectively.
Next, we can ask for the 3D diffeomorphisms preserving this form above. We write it as
\begin{align}\label{PBHY1}
x^{\mu}\rightarrow &x^{\mu}-a^{(1)}[e^{2 s} y,x] \partial^{\mu} \tau+O\left(\partial^3\right),\\\label{PBHY2}
y \rightarrow & e^{2 s} y+b^{(3)}[e^{2 s} y,x] (\partial \tau)^2+O\left(\partial^4\right),
\end{align}
where the covariant form factors obey the following constraints
\begin{align}
\partial_z a^{(1)}[z,x]=2\frac{g_{yy}(z)}{z^2}, \ \  \partial_z b^{(1)}+\left(\frac{\partial_z g_{yy}}{2 g_{yy}}- \frac{1}{z} \right) b^{(1)}+\frac{z^3}{2 g_{yy}} (\partial_z a^{(1)})^2=0,
\end{align}
which can be solved easily for a given RG flow metric in this gauge.
\subsection{Near To Boundary Analysis}
We use the near to boundary analysis to reproduce the results for the bulk action in presence of a PBH mode and to compute the GH and counterterm contribution. We start by writing the near to boundary expansion of the equations of motion. We then evaluate the onshell bulk contribution and finally the onshell contributions from GH and counterterm. 
\label{sec:EoMApp}
 \renewcommand{\arraystretch}{1.2}
 \enlargethispage{\baselineskip}
\subsubsection{Near to boundary expansion of the EoM} \label{NTBExpansion}

The near to boundary expansion of the equations of motion in the Fefferman-Graham gauge choice (\ref{metric}) comes from:
\begin{eqnarray*}\label{NearToBEoM}
y \lbrack 2g^{\prime \prime }_{ij}-2(g^{\prime }g^{-1}g^{\prime
})_{ij}+Tr(g^{-1}g^{\prime })g_{ij}]+R_{ij}-2g_{ij}-Tr(g^{-1}g^{\prime })g_{ij} &=&%
\frac{4}{3}\frac{g_{ij}}{y }(V[\phi ]-V_{fp}) \\
Tr(g^{-1}g^{\prime \prime })-\frac{1}{2}Tr(g^{-1}g^{\prime }g^{-1}g^{\prime
}) &=&\frac{8}{3}g_{y y }(V[\phi ]-V_{fp})+8(\phi ^{\prime })^{2},
\end{eqnarray*}
where the primes denote derivative with respect to the flow variable $y$ and $V_{fp}$ is the potential at the corresponding fixed point. For the case of boundary $V_{fp}=V[0]$.
 
 Another useful relation that is going to be helpful in computing the spurion effective action is the following form for the onshell action:
\begin{align}\label{onshellLag}
S_{bulk}^{os}=\frac{L_{UV}}{2}\int d^{4}x\int dy \sqrt{g}(-\frac{2}{3}V[\phi ]).
\end{align}

\paragraph{Solutions}
We can solve the equations of motions for a generic potential of the form (\ref{Potential}). Let us start by the UV side.
\paragraph{The UV side}
We can check now the result (\ref{bulkEffectiveAction}), for the bulk action after a $\tau$:PBH (except for the finite part of course) by using near to boundary analysis. As said before we take the near to boundary expansion of the scalar field to be:
\[
\phi \sim y \phi ^{(0)}(x)+y \log (y )\widetilde{\phi }^{(0)}(x),
\]
where the $\widetilde{\phi }^{(0)}$ and $\phi ^{(0)}$ are identified with the source and vev of
a dimension $\Delta=2$ CFT operator, respectively.
 The terms in the near to boundary expansion  (\ref{NToBExpansion1}) of the metric are solved to be:
\[
g_{ij}^{(2)}=\frac{1}{2}(R_{ij}[g^{(0)}]-\frac{1}{6}g_{ij}^{(0)}R),
\]
\begin{eqnarray}
Tr(h_{1}^{(4)}) &=&\frac{16}{3}\phi _{(0)}\widetilde{\phi }_{(0)},\label{h1} \\
Tr(h_{2}^{(4)}) &=&\frac{8}{3}\widetilde{\phi }_{(0)}^{2},\label{GHcoeff0} \\
Tr(g^{(4)}) &=&\frac{1}{4}tr(g_{(2)}^{2})-\frac{3}{2}%
tr(h_{1}^{(4)})-tr(h_{2}^{(4)})+\frac{8}{3}\phi _{(0)}^{2}+4\widetilde{\phi }%
_{(0)}^{2}+8\phi _{(0)}\widetilde{\phi }_{(0)}\\
&=&\frac{1}{4}tr(g_{(2)}^{2})+\frac{8}{3}\phi _{(0)}^{2}+\frac{16}{3}%
\widetilde{\phi }_{(0)}^{2}.
\end{eqnarray}
The volume measure expansion:
\begin{eqnarray}
\sqrt{g} &=&\sqrt{g_{(0)}}(1+\frac{1}{2}Tr(g_{(2)})y +(\frac{1%
}{2}Tr(g_{(4)})+\frac{1}{8}Tr(g_{(2)})^{2}-\frac{1}{4}Tr(g_{(2)}^{2})\nonumber \\
&&+\frac{1}{2}Tr(h_{(4)}^{1})\log (y )+\frac{1}{2}Tr(h_{(4)}^{2})\log
^{2}(y ))y ^{2}),\label{volumeFactor}
\end{eqnarray}
is used to evaluate the near to boundary expansion of bulk lagrangian in (\ref{onshellLag}). The result for
the UV expansion of the onshell action (\ref{NTBAction}), is evaluated by use of the following result for a conformally flat $g_{(0)}=e^{-\tau}\eta$
\begin{eqnarray}
a_{UV}^{(0)} &=&\frac{1}{2 L_{UV}}\int d^{4}x \sqrt{g_{(0)}}=\frac{1}{2 L_{UV}}\int d^{4}x e^{-2\tau },\nonumber \\
a_{UV}^{(2)} &=&\frac{L_{UV}}{2}\int d^{4}x\sqrt{g_{(0)}}Tr(g_{(2)})=\frac{L_{UV}}{8}\int d^{4}x e^{-\tau }(\partial \tau)^2, \label{as} \\
a_{UV}^{(4)} &=&L_{UV}^{3}\int d^{4}x\sqrt{g_{(0)}}(\frac{1}{2}%
Tr(g_{(4)})+\frac{1}{8}Tr(g_{(2)})^{2}-\frac{1}{4}Tr(g_{(2)}^{2})-\frac{4}{3}%
\phi _{(0)}^{2})=L_{UV}^{3}\int d^{4}x\frac{8}{3}\tilde{\phi}_{(0)}^2.\nonumber
\end{eqnarray}
Use of Weyl transformations properties of the Ricci scalar in 4D was used in getting this result.
\paragraph{GH term contribution}
\label{AppGH}
In the UV side we can expand the Gibbons Hawking term in a near to boundary series:
\begin{eqnarray}
\frac{1}{4}\int d^{4}x\sqrt{\gamma }2K|_{UV}=\frac{1}{L_{UV}}\int d^{4}x\frac{1}{y_{UV}^2}(-2\sqrt{g}%
+y \partial _{y }\sqrt{g}) \nonumber\\
=\int d^4 x\left( \frac{b^{(0)}}{y _{UV}^{2}}+\frac{b^{(2)}}{y
_{UV}}+b^{(4)}\log (y_{UV})+b_{finite}\right),
\end{eqnarray}
where,
\begin{gather}
b^{(0)}=-\frac{2}{L_{UV}}\int d^{4}x\sqrt{g_{(0)}},  \ b^{(2)}=-\frac{L_{UV}}{2}\int d^{4}x\sqrt{g_{(0)}}Tr(g_{(2)}),
\\
b^{(4)}=L_{UV}^{3}\int d^{4}x\sqrt{g_{(0)}}Tr(h_{2}^{(4)}), \
b_{finite}=\frac{L_{UV}^{3}}{2}\int d^{4}x\sqrt{g_{(0)}}Tr(h_{1}^{(4)}).\label{GHcoeff1}
\end{gather}
The finite contribution $b_{finite}$ is proportional to $ \int d^4 x \ Tr(h_{1}^{(4)})$
which by (\ref{h1}) is proportional to the product of the vev and the source $\phi^{(0)}$ and $\tilde{\phi}^{(0)}$ respectively.
Namely, for a vev driven flow the GH term does not contribute at all to the finite part of the regularized onshell action.
In the case of a source driven flow, the finite contribution gives a potential term which is not Weyl invariant, as one can notice from the transformation properties (\ref{vev}).  In fact its infinitesimal Weyl transformation generates an anomalous variation proportional to the source square $\delta \tau(\tilde{\phi}^{(0)})^2$. This fact can be notices by simple eye inspection one just need to analyse the transformations
properties (\ref{vev}) on the static case.
\paragraph{The IR side}
In this case we can do the same. As already said, we assume IR regularity in the corresponding background, namely,
\[
\phi \sim \phi_{IR}+ \frac{1}{\rho ^{m}}\phi ^{(0)}+..., \ m > 0.
\]
We start by writing the IR asymptotic expansion of the GH term in the IR:
\[
\frac{1}{4}\int d^{4}x\sqrt{\gamma }2K|_{IR}\sim\int d^4x\left( \frac{b_{IR}^{(0)}}{y _{IR}^{2}}+\frac{b_{IR}^{(2)}}{y _{IR}}%
+b_{IR}^{(4)}\log y +b_{finite}+\sum_{n=1}^{\infty }y _{IR}^{n}b_{IR}^{(n)}\right).
\]
We compute the factors  $b$ in terms of the components of the near to IR expansion of the metric:
\begin{gather}
b_{IR}^{(0)} =\frac{1}{2l_{IR}}\int d^{4}x\sqrt{g_{(0)}} , \
b_{IR}^{(2)} =\frac{L_{IR}}{2}\int d^{4}x\sqrt{g_{(0)}}e^{-2\tau
}Tr(g_{(2)}), \\
b_{IR}^{(4)}=L_{IR}^{3}\int d^{4}x\sqrt{g_{(0)}}e^{-2\tau }(\frac{1}{2}
Tr(g_{(4)})+\frac{1}{8}Tr(g_{(2)})^{2}-\frac{1}{4}Tr(g_{(2)}^{2})), \\
b_{finite} = L_{IR}^{3}\int d^{4}x\sqrt{g_{(0)}}e^{-2\tau }Tr(h_{1}^{(4)}).
\end{gather}
By using the near to IR expansion of the equations of motions (\ref{NearToBEoM}) at second order we get:
\[
g_{ij}^{(2)}=\frac{1}{2}(R_{ij}[g^{(0)}]-\frac{1}{6}g_{ij}^{(0)}R),
\]
and additionally:
\begin{gather*}
Tr(h_{1}^{(4)}) =0, \
Tr(h_{2}^{(4)}) =0, \\
Tr(g^{(4)}) =\frac{1}{4}tr(g_{(2)}^{2})-\frac{3}{2}
tr(h_{1}^{(4)})-tr(h_{2}^{(4)})=\frac{1}{4}tr(g_{(2)}^{2}).
\end{gather*}
It is then easy to see how the IR GH term does not contribute to the finite part of the regularized action! provided the background solutions are smooth in the IR.


\subsection{Anomaly matching from PBH transformations} \label{AnomalyM}
In this appendix we present an alternative way to compute the gravitational WZ term. The approach is covariant in the sense that It works with an arbitrary boundary background metric $g^{(0)}$ and shows how the 4D anomaly matching argument of \citep{Komargodski1,Schwimmer1}  is linked to the 5D PBH transformation properties.

The relevant terms in the cut off expansion of the bulk action are:
\begin{gather}\label{ac}\nonumber
\hspace{-5cm}S[\tau]=\int d^4 x\sqrt{\hat{g}^{0}} (\frac{1}{y_{UV}^2}-\frac{1}{y_{IR}^2}+\frac{a^{(2)}_{UV}[\hat{g}^{(0)},\hat{\phi}^{(0)}]}{y_{UV}}-\frac{a^{(2)}_{IR}[\hat{g}^{(0)},\hat{\phi}^{(0)}]}{y_{IR}}\\ \hspace{4 cm} +a^{(4)}_{UV}[\hat{g}^{(0)},\hat{\phi}^{(0)}]\log(y_{UV})-a^{(4)}_{IR}[\hat{g}^{(0)},\hat{\phi}^{(0)}]\log(y_{IR}) )+S_{finite}[\tau]+\ldots,  \end{gather}
after a finite PBH transformation parameterized by $\tau$ is performed. The $S_{finite}[\tau]$ stands for the cut off independent contribution to the bulk action and $\hat{g}^{(0)}=e^{-\tau} g^{(0)}$ and $\hat{\phi}^{(0)}$ stand for the PBH transformed boundary data. The leading "matter" boundary data $\hat{\phi}^{(0)}$ (UV/IR need not be the same), do not  transform covariantly,  unlike the background boundary metric $g^{(0)}$. 

Next, one can perform a second infinitesimal PBH, $\delta \tau_1$, and think about it in two different ways:
\begin{itemize}
\item Keep the cut-off fixed and transform the fields (I).
\item Keep the fields fixed and transform  the cut-offs (II).
\end{itemize}

In approach $I$, in virtue of additivity of PBH transformations:

\begin{equation}  \delta S_{finite}= \delta \tau_1\left( \frac{\delta S_{finite}[\tau]}{\delta \tau}\right). \label{A1}\end{equation}

In approach $II$, one needs the generalization of (\ref{PBH}) for a linear parameter $\delta \tau_1$ and arbitrary boundary metric $g^{(0)}$. An important point is that (\ref{PBH}) is not a near to boundary expansion, but rather an IR expansion valid along the full flow geometry. Notice also that, in principle, some contribution proportional to $\Box \delta \tau_1$, $\Box \Box \delta \tau_1$, .., could come out of the cut off powers in (\ref{ac}). As discussed for (\ref{ac}), these terms can be completely gauged away. Then approach $II$ gives:

\begin{equation}\delta S_{finite}=\int d^4 x \sqrt{\hat{g}^{0}}\ \delta \tau_1 \left( a^{(4)}_{UV}[\hat{g}^{(0)},\hat{\phi}^{(0)}]-a^{(4)}_{IR}[\hat{g}^{(0)},\hat{\phi}^{(0)}]  \right).\label{A2}\end{equation} 
Equating (\ref{A1}) and (\ref{A2}) we get:
\begin{equation}
 \frac{\delta S_{finite}[\tau]}{\delta \tau}=\int d^4 x \sqrt{\hat{g}^{0}}\ \left( a^{(4)}_{UV}[\hat{g}^{(0)},\hat{\phi}^{(0)}]-a^{(4)}_{IR}[\hat{g}^{(0)},\hat{\phi}^{(0)}]  \right).\label{WZeval}
\end{equation}
Now we can expand the gravitational contribution to $a^{(4)}_{UV}[\hat{g}^{(0)}]-a^{(4)}_{IR}[\hat{g}^{(0)}]$: \[ \frac{\left(L_{UV}^3-L_{IR}^3\right)}{64} \left(E_{(4)}[\hat{g}^{(0)}]-W[\hat{g}^{(0)}]^2\right),\]  by using the Weyl expansions:
\begin{eqnarray}
\hat{W}^2 &= &e^{2 \tau}W^2,\nonumber
\\
\hat{E}_{(4)}&= &e^{2 \tau} \left(E_{(4)} +4\left(R^{\mu\nu} -\frac{ 1}{2}
{g^{(0)}}^{\mu\nu}R\right)\nabla_{\mu}\partial_{\nu}\tau\right) \nonumber\\&& \  \ ~~~~~~~~ 
+e^{2 \tau}\left( 2\left((\Box \tau )^2 - \Box^{\mu\nu}\tau \Box_{ \mu \nu}\tau\right)- \left((\Box \tau)(\partial \tau)^2 +2\ \partial_{\mu}\tau \Box^{\mu \nu}\tau \partial_{\nu} \tau\right) \right).\nonumber\\ \label{Trans}
\end{eqnarray}
Hence, from (\ref{WZeval}) and (\ref{Trans}) one can integrate out the gavitational contribution to $S_{finite}$:
\begin{equation}
 \int d^4 x\sqrt{g^{0}} \ \left( \Delta a \left(E_{(4)}\frac{\tau}{2} -\left(R^{\mu\nu} -\frac{ 1}{2}
{g^{(0)}}^{\mu\nu}R\right)\partial_{\mu}\tau\partial_{\nu}\tau+\frac{1}{8}\left((\partial \tau)^4-4\ \Box \tau (\partial \tau)^2\right) \right)-\Delta c \ W^2 \frac{\tau}{2}\right),
\end{equation}
where in the case we are considering $c= a$. Notice that in the above derivatiom, 
we implicitly assumed the group property of the PBH transformations on fields, that is:
\[L_{\tau_{1}}\circ L_{\tau_{2}}=L_{\tau_1+\tau_2},\] were $L$ represents the transformation thought of as an operator acting on the fields (boundary data).  As for the case of matter contributions, a problem arises when a v.e.v. or source transforms non covariantly  \[\phi^{(0)} \rightarrow e^{\tau} \phi^{(0)}+\tau e^{\tau} \tilde{\phi}^{(0)}.\]
So, it is not clear to us how to use this procedure to compute "matter" contributions to the Weyl anomaly. An efficient procedure to compute anomalies for generic backgrounds (in a spirit similar to the approach presented here), had appeared in \citep{Schwimmer3} (section \S 3.1).

\section{3D N=4 SUGRA example}

\subsection{Equations of motion for the background fluctuations}\label{EqFluctuations}
We start by writing down the gravitational side of the set of equations of motion for the background fluctuations $g^{(2)}$, $T^{(2)}$, $g_{tx}^{(2)}$, $A^{(2)}$ and $\phi^{(2)}$ at second order in time $t$ and space $x$ derivatives. We use here the notation used through out the main text, namely denoting the equations as the space time components they descend from. So the equations $(r,r)$, $(t,t)-(x,x)$, $(t,t)+(x,x)$ and $(t,x)$, read off respectively:

\begin{multline}\partial^2_r g^{(2)}+2 \partial_r f_B \partial_r g^{(2)} +4\partial_{\phi_B} V \phi^{(2)}+2\partial_r \phi_B \partial_r\phi^{(2)}\\+4\left(\partial_{A_B} V + \frac{3 A_B}{(1-A_B^2)^3} (\partial_r A_B)^2 \right)A^{(2)}+ \frac{6}{(1-A_B^2)^2}\partial_r A_B\partial_r A^{(2)}=0, \end{multline}
\begin{multline}\partial^2_r g^{(2)}+4 \partial_r f_B \partial_r g^{(2)} +2\left( 4 V+2 (\partial_r f_B)^2+\partial^{2}_r f_B \right) g^{(2)}\\+8\partial_{\phi_B} V \phi^{(2)}+8\partial_{A_B} V A^{(2)}+e^{-2 f_B}\left(\frac{3}{(1-A_B^2)^2}(\partial A_B)^2+(\partial \phi_B)^2+2 \Box f_B \right)=0, \end{multline}
\begin{multline}\partial^2_r T+2 \partial_r f_B \partial_r T +2\left(4 V +2(\partial_r f_B)^2+\partial^2_r f_B \right)T \\ -e^{-2 f_B}\left(\frac{3}{(1-A_B^2)^2}(\partial A_B)^2+(\partial \phi_{B})^2\right)=0, \end{multline}
\begin{multline}\partial^2_r g^{(2)}_{tx}+2 \partial_r f_B \partial_r g^{(2)}_{tx} +2\left(4 V +2(\partial_r f_B)^2+\partial^2_r f_B \right)g^{(2)}_{tx} \\ +2 e^{-2 f_B}\left(\frac{3}{(1-A_B^2)^2}(\partial A_B)^2+(\partial \phi_{B})^2\right)=0, \end{multline}
where for a  $Y\equiv A_B,\phi_B,f_B$, we use the notation $(\partial Y)^2\equiv(\partial_x Y)^2-(\partial_t Y)^2$ and $\Box Y=(\partial^2_x Y-\partial^2_t Y)$. We also used the equations $(t,r)$ and $(x,r)$ respectively:

\begin{multline}
\left(\partial^2_{t r}T +2\partial_t f_B \partial_r T \right)+\left(\partial^2_{x r}g^{(2)}_{t x}+2\partial_x f_B \partial_r g^{(2)}_{t x}\right)-\partial^2_{t r}g^{(2)}-2\left(\partial_t \phi_B \partial_r\phi^{(2)}+\partial_r\phi_B \partial_t\phi^{(2)}\right)\\-\left(\frac{6}{(1-A_B^2)^2}\left(\partial_t A_B \partial_r A^{(2)}+\partial_r A_B \partial_t A^{(2)}\right)+\frac{24}{(1-A_B^2)^3}(A_B \partial_r A_B \partial_t A_B)  A^{(2)}\right)=0,
\end{multline}
\begin{multline}
-\left(\partial^2_{x r}T +2\partial_x f_B \partial_r T \right)-\left(\partial^2_{t r}g^{(2)}_{t x}+2\partial_t f_B \partial_r g^{(2)}_{t x}\right)-\partial^2_{x r}g^{(2)}-2\left(\partial_x \phi_B \partial_r\phi^{(2)}+\partial_r\phi_B \partial_x\phi^{(2)}\right)\\-\left(\frac{6}{(1-A_B^2)^2}\left(\partial_x A_B \partial_r A^{(2)}+\partial_r A_B \partial_x A^{(2)}\right)+\frac{24}{(1-A_B^2)^3}(A_B \partial_r A_B \partial_x A_B)  A^{(2)}\right)=0.
\end{multline}
These equations reduced to constraints for the integration constants that appear.

The Klein-Gordon equations for the scalar fields $\phi$ and $A$ give the following couple of equations for the fluctuations respectively:
\begin{align}
\partial_r^2 \phi^{(2)}+2 \partial_r f_B \partial_r \phi^{(2)}-2 \partial^2_{\phi_B} V \phi^{(2)}-2 \partial^2_{A_B,\phi_B}V A^{(2)}+\partial_r \phi_B \partial_r g^{(2)}+e^{-2 f_B} \Box \phi_B=0,
\end{align}
\begin{multline}
\partial_r^2 A^{(2)}+\left(2 \partial_r f_B+\frac{4 A_B}{(1-A_B^2)^2}\partial_r A_B \right)\partial_r A^{(2)}\\+\frac{2}{3}\left(-(1-A_B^2)^2 \partial^2_{A_B}V+3\frac{(1-5A_B^2)}{(1-A_B^2)^2}(\partial_r A_B)^2+\frac{6A_B}{(1-A_B^2)}\left(2\partial_r A_B \partial_r f_B+\partial^2_r A_B\right) \right)A^{(2)}\\
+\partial_r A_B \partial_r g^{(2)}+\frac{2}{3}(1-A_B^2)^2\partial^2_{A_B,\phi_B}V \phi^{(2)}+e^{-2f_B}\left(\frac{A_B}{1-A_B^2}(\partial A_B)^2 +\Box A_B\right)=0.
\end{multline}

\subsection{Rational functions for the pair $(s_{p},\tau)$}
\label{rationalFunctions}
In this subsection we write down the rational functions appearing in the equations in section \S 3.

\begin{align}\nonumber
R^{(1)}_{\partial_y \phi ^{(2)}}= -\frac{3 g_2^3 (y+1)^3  }{g_1 (g_1^2-g_2^2 (y+1)^2)},& \
R^{(2)}_{\partial_y \phi ^{(2)}}=-\frac{ \left(g_1^2 (2 y+1)+g_2^2 (y+1)^2 (2 y-1)\right)}{y (y+1) \left(g_2^2 (y+1)^2-g_1^2\right)},
\\\nonumber
R^{(3)}_{\partial_y \phi ^{(2)}}=\frac{3 g_2^3 (y+1)^3 \left(g_2^2 \left(y^2-1\right)+g_1^2\right)}{g_1 y \left(g_1^2-g_2^2 (y+1)^2\right){}^2},& \
R^{(4)}_{\partial_y \phi ^{(2)}}=\frac{(g_2^2 (y+1)^3+g_1^2 (y-1))}{2 g_1^2 y^3}, \\
R^{(5)}_{\partial_y \phi ^{(2)}}=\frac{(y+1) (3 g_2^2 (y+1)^2+g_1^2)}{y(g_1^2-g_2^2 (y+1)^2){}^2}, & \
R^{(6)}_{\partial_y \phi ^{(2)}}=-\frac{2 c_1^2 \left(g_1^2-g_2^2 (y+1)^2\right){}^2}{g_1^6 g_2^4 y^3 (y+1)},
\end{align}

\begin{align}
F^{(1)}&=-\frac{2 \left(g_1^4 \left(5 y^2+6 y+2\right)+2 g_2^4 (y+1)^5-4 g_1^2 g_2^2 (y+1)^4\right)}{g_1 g_2 y^4 (y+1)^3 \left(g_1^2 (3 y+2)-2
g_2^2 (y+1)^2\right)}e^{2 sp},
\\\nonumber
F^{(2)}&=\frac{4 c_1^2 \left(g_1^2 (3 y+2)-2 g_2^2 (y+1)^2\right) \left(g_1^2-g_2^2 (y+1)^2\right)}{g_1^5 g_2^5 y^4 (y+1)^4},
\\\nonumber
F^{(3)}&=-\frac{4 c_1 \left(g_1^4 \left(5 y^2+6 y+2\right)+2 g_2^4 (y+1)^5-4 g_1^2 g_2^2 (y+1)^4\right)}{g_1^3 g_2^3 y^4 (y+1)^3 \left(g_1^2 (3
y+2)-2 g_2^2 (y+1)^2\right)},\\ \nonumber
F^{(4)}&=\frac{2 g_1 \left(g_1^6
   \left(12 y^2+13 y+4\right)-4 g_2^6 (y+1)^8\right)}{g_2 y^3 (y+1)^2 \left(g_2^2 (y+1)^2-g_1^2\right){}^3 \left(2 g_2^2 (y+1)^2-g_1^2 (3
   y+2)\right)}\\ & \qquad +\frac{2 g_1^2 g_2^4 (y+1)^4 \left(9 y^3+32 y^2+29 y+12\right)-2 g_1^4 g_2^2 (y+1)^2 \left(21 y^3+40 y^2+34
   y+12\right)}{g_2 y^3 (y+1)^2 \left(g_2^2 (y+1)^2-g_1^2\right){}^3 \left(2 g_2^2 (y+1)^2-g_1^2 (3 y+2)\right)},
\end{align}

\begin{multline}
R^{(0)}_{A^{(2)}}  =\frac{8 g_2^8 (y+1)^8-2 g_1^2 g_2^6 (y+1)^5 (y (y (8 y+27)+29)+16)}{y^3 (y+1)^2 \left(g_2^2
   (y+1)^2-g_1^2\right){}^3 \left(2 g_2^2 (y+1)^2-g_1^2 (3 y+2)\right)}\\ + \frac{2 \left(g_1^8 (y (12 y+13)+4)+g_2^4 g_1^4 (y+1)^3 (y (y (2 y (9 y+32)+79)+63)+24)\right)}{y^3 (y+1)^2
\left(g_2^2 (y+1)^2-g_1^2\right){}^3 \left(2 g_2^2 (y+1)^2-g_1^2 (3 y+2)\right)} \\-\frac{2 g_1^6 g_2^2 (y+1) \left(y
\left(y \left(42
   y^2+68 y+77\right)+55\right)+16\right)}{y^3 (y+1)^2 \left(g_2^2
   (y+1)^2-g_1^2\right){}^3 \left(2 g_2^2 (y+1)^2-g_1^2 (3 y+2)\right)},
\end{multline}
\begin{multline}
R^{(1)}_{A^{(2)}}=\frac{4 g_1^4 g_2^2 (y+1)^2 (y (y+1) (15 y-13)-6)-2 g_1^2 g_2^4 (y+1)^4 (y (2 y (9 y+13)-13)-12)}{y^2 (y+1)^2 \left(g_1^2-g_2^2
(y+1)^2\right){}^2 \left(2 g_2^2 (y+1)^2-g_1^2 (3 y+2)\right)}
\\
+\frac{2 \left(2 g_2^6 (y+1)^6 \left(5
   y^2-2\right)+g_1^6 (y (12 y+13)+4)\right)}{y^2 (y+1)^2 \left(g_1^2-g_2^2 (y+1)^2\right){}^2 \left(2 g_2^2 (y+1)^2-g_1^2 (3 y+2)\right)},
\end{multline}

\begin{equation}
R^{(2)}_{A^{(2)}} =\frac{g_2}{g_2
y-g_1+g_2}+\frac{g_2}{g_2 y+g_1+g_2}+\frac{4 g_2^2 (y+1)-3 g_1^2}{g_1^2 (3 y+2)-2 g_2^2 (y+1)^2}+\frac{2}{y}+\frac{6}{y+1} ,
\end{equation}

\begin{gather}
l^{(0)}=\frac{g_1^2 g_2^2 y \left(g_2^2 (y+1)^3+g_1^2 (y-1)\right) }{8 c_1 (y+1) \left(g_2^2 (y+1)^2-g_1^2\right)}e^{2 sp}, \
l^{(1)}=\frac{g_1^4 g_2^2 y \left(3 g_2^2 (y+1)^2+g_1^2\right)}{4 c_1 \left(g_1^2-g_2^2 (y+1)^2\right){}^3}, \\
l^{(2)}=\frac{g_2^2 (y+1)^3+g_1^2 (y-1)}{4 y (y+1) \left(g_2^2 (y+1)^2-g_1^2\right)}, \
l^{(3)}=-\frac{c_1 \left(g_1^2-g_2^2 (y+1)^2\right)}{2 g_1^2 g_2^2 y (y+1)^2}, \\
l^{(4)}=\frac{g_1^2 g_2^2y^2}{4 c_1}e^{2sp}, \
l^{(5)}=\frac{g_1^2 g_2^2 y \left(g_2^2 y^3+3 g_2^2 y^2+\left(g_1^2+3 g_2^2\right) y-g_1^2+g_2^2\right)}{8 c_1 (y+1) \left(g_2^2 y^2+2 g_2^2
y-g_1^2+g_2^2\right)}e^{2sp}, \\
l^{(6)}=-\frac{3 g_1^3 g_2^5 e^{2 \text{sp}} y^2 (y+1)^2}{4 c_1 \left(g_2^2 y^2+2 g_2^2 y-g_1^2+g_2^2\right){}^2}, \
l^{(7)}= \frac{g_1^4 g_2^2 e^{2 \text{sp}} y^2}{4 c_1 (y+1) \left(g_2^2 y^2+2 g_2^2 y-g_1^2+g_2^2\right)}.
\end{gather}
\paragraph{Case of the modulus $\rho$}

\begin{gather}\nonumber
R^{(1)}_{\partial_y \phi ^{(2)}}= -\frac{3 g_2^3 (y+\rho)^3  }{\rho g_1 (g_1^2-g_2^2 (y+\rho)^2)}, \
R^{(2)}_{\partial_y \phi ^{(2)}}=-\frac{g_1^2 \rho ^2 (\rho +2 y)+g_2^2 (2 y-\rho ) (\rho +y)^2}{y (\rho +y) \left(g_2^2 (\rho +y)^2-g_1^2 \rho ^2\right)},
\\\nonumber
R^{(3)}_{\partial_y \phi ^{(2)}}=\frac{3 g_2^3 (\rho +y)^3 \left(g_1^2 \rho ^2+g_2^2 \left(y^2-\rho ^2\right)\right)}{g_1 \rho  y \left(g_1^2 \rho ^2-g_2^2 (\rho +y)^2\right){}^2}, R^{(4)}_{\partial_y \phi ^{(2)}}=\frac{g_1^2 \rho ^2 (y-\rho )+g_2^2 (\rho +y)^3}{2 g_1^2 \rho ^2 y^3} ,\\
R^{(5)}_{\partial_y \phi ^{(2)}}=\frac{c_1^2 \left(g_1^2 \rho ^2-g_2^2 \left(\rho ^2+3 y^2+4 \rho  y\right)\right)}{g_1^4 g_2^4 \rho  y^3 (\rho +y)}, \end{gather}
\begin{multline}
R^{(6)}_{\partial_y \phi ^{(2)}}=\frac{c_1^2 \left(g_1^6 \rho ^5-g_2^4 g_1^2 \rho  (\rho +y)^2 \left(3 \rho ^2+8 y^2+10 \rho  y\right)\right)}{g_1^4 g_2^4 \rho  y^3
   (\rho +y) \left(g_1^2 \rho ^2-g_2^2 (\rho +y)^2\right){}^2}\\+\frac{c_1^2\left(g_2^2 g_1^4 \rho ^3 \left(3 \rho ^2+5 y^2+8 \rho  y\right)+g_2^6 (\rho +y)^4 (\rho +4 y)\right)}{g_1^4 g_2^4 \rho  y^3
   (\rho +y) \left(g_1^2 \rho ^2-g_2^2 (\rho +y)^2\right){}^2},
\end{multline}

\begin{multline}
F^{(1)}=\frac{2 c_1^2 \left(-g_1^2 g_2^2 \rho  \left(8 \rho ^4+9 y^4+32 \rho  y^3+47 \rho ^2 y^2+32 \rho ^3 y\right)+g_1^4 \rho ^3 (2 \rho +3 y)^2+4 g_2^4 (\rho +y)^5\right)}{g_1^3 g_2^5 y^4 (\rho +y)^4 \left(2 g_2^2 (\rho +y)^2-g_1^2 \rho  (2 \rho
   +3 y)\right)},
\end{multline}
\begin{multline}
F^{(2)}=-\frac{2 c_1^2 \left(-2 g_2^8 g_1^2 \rho  (\rho +y)^6 \left(10 \rho ^3+6 y^3+27 \rho  y^2+28 \rho ^2 y\right)\right)}{g_1^3 g_2^5 y^4 (\rho +y)^4 \left(g_2 (\rho +y)-g_1 \rho \right)^3 \left(g_1 \rho +g_2 (\rho +y)\right)^3 \left(2 g_2^2 (\rho +y)^2-g_1^2 \rho  (2 \rho +3 y)\right)}\\-\frac{2 c_1^2 \left(g_2^4 g_1^6 \rho ^4 (\rho +y)^2 \left(40 \rho ^4+54 y^4+178 \rho  y^3+251 \rho ^2 y^2+164 \rho ^3 y\right)\right)}{g_1^3 g_2^5 y^4 (\rho +y)^4 \left(g_2 (\rho +y)-g_1 \rho \right)^3 \left(g_1 \rho +g_2 (\rho +y)\right){}^3 \left(2 g_2^2 (\rho +y)^2-g_1^2 \rho  (2 \rho +3 y)\right)}\\+\frac{2 c_1^2 \left(g_2^6 g_1^4 \rho ^2
   (\rho +y)^4 \left(40 \rho ^4+15 y^4+98 \rho  y^3+180 \rho ^2 y^2+140 \rho ^3 y\right)\right)}{g_1^3 g_2^5 y^4 (\rho +y)^4 \left(g_2 (\rho +y)-g_1 \rho \right)^3 \left(g_1 \rho +g_2 (\rho +y)\right)^3 \left(2 g_2^2 (\rho +y)^2-g_1^2 \rho  (2 \rho +3 y)\right)}\\+\frac{2 c_1^2 \left(2 g_2^2 g_1^8 \rho ^6 \left(10 \rho ^4+18 y^4+61 \rho  y^3+79 \rho^2 y^2+46 \rho ^3 y\right)\right)}{g_1^3 g_2^5 y^4 (\rho +y)^4 \left(g_2 (\rho +y)-g_1 \rho \right)^3 \left(g_1 \rho +g_2 (\rho +y)\right)^3 \left(2 g_2^2 (\rho +y)^2-g_1^2 \rho  (2 \rho +3 y)\right)}\\-\frac{2 c_1^2 \left(g_1^{10}\rho ^8 (2 \rho +3 y)^2+4 g_2^{10}
   (\rho +y)^{10}\right)}{g_1^3 g_2^5 y^4 (\rho +y)^4 \left(g_2 (\rho +y)-g_1 \rho \right)^3 \left(g_1 \rho +g_2 (\rho +y)\right)^3 \left(2 g_2^2 (\rho +y)^2-g_1^2 \rho  (2 \rho +3 y)\right)},
\end{multline}
\begin{equation}
F^{(3)}=-\frac{2 \left(g_1^4 \rho ^3 \left(2 \rho ^2+5 y^2+6 \rho  y\right)+2 g_2^4 (\rho +y)^5-4 g_1^2 g_2^2 \rho  (\rho +y)^4\right)}{g_1 g_2 y^4 (\rho +y)^3 \left(2 g_2^2 (\rho +y)^2-g_1^2 \rho  (2 \rho +3 y)\right)},
\end{equation}

\begin{multline}
R^{(0)}_{A^{(2)}}=\frac{2 g_2^2 \rho ^2 y \left(g_2^2 g_1^4 \rho  \left(510 \rho ^3+18 y^3+118 \rho  y^2+325 \rho ^2 y\right)\right)}{(\rho +y)^2 \left(g_2^2 (\rho +y)^2-g_1^2 \rho ^2\right){}^3 \left(2 g_2^2 (\rho +y)^2-g_1^2 \rho  (2 \rho +3 y)\right)}\\+\frac{2 g_2^2 \rho ^2 y \left(+4 g_2^6 \left(70 \rho ^4+y^4+8 \rho  y^3+28
   \rho ^2 y^2+56 \rho ^3 y\right)-2 g_1^6 \rho ^3 (55 \rho +21 y)\right)}{(\rho +y)^2 \left(g_2^2 (\rho +y)^2-g_1^2 \rho ^2\right){}^3 \left(2 g_2^2 (\rho +y)^2-g_1^2 \rho  (2 \rho +3 y)\right)}\\+\frac{2 g_2^2 \rho ^2 y \left(-g_2^4 g_1^2 \left(680 \rho ^4+8 y^4+67 \rho  y^3+244 \rho ^2 y^2+511 \rho ^3 y\right)\right)}{(\rho +y)^2 \left(g_2^2 (\rho +y)^2-g_1^2 \rho ^2\right){}^3 \left(2 g_2^2 (\rho +y)^2-g_1^2 \rho  (2 \rho +3 y)\right)},
\end{multline}

\begin{multline}
R^{(1)}_{A^{(2)}}=\frac{2 \left(2 g_2^6 (\rho +y)^6 \left(5 y^2-2 \rho ^2\right)+g_1^6 \rho ^6 \left(4 \rho ^2+12 y^2+13 \rho  y\right)\right)}{y^2 (\rho +y)^2 \left(g_1^2 \rho ^2-g_2^2 (\rho +y)^2\right){}^2 \left(2 g_2^2 (\rho +y)^2-g_1^2 \rho  (2 \rho +3 y)\right)}\\+\frac{2 \left(+2 g_2^2 g_1^4 \rho ^3 (\rho +y)^2 \left(-6 \rho ^3+15 y^3+2 \rho  y^2-13 \rho ^2 y\right)\right)}{y^2 (\rho +y)^2 \left(g_1^2 \rho ^2-g_2^2 (\rho +y)^2\right){}^2 \left(2 g_2^2 (\rho +y)^2-g_1^2 \rho  (2 \rho +3 y)\right)}\\+\frac{2\left(g_2^4 g_1^2 \rho  (\rho +y)^4
   \left(12 \rho ^3-18 y^3-26 \rho  y^2+13 \rho ^2 y\right)\right)}{y^2 (\rho +y)^2 \left(g_1^2 \rho ^2-g_2^2 (\rho +y)^2\right){}^2 \left(2 g_2^2 (\rho +y)^2-g_1^2 \rho  (2 \rho +3 y)\right)},
\end{multline}
\begin{equation}
R^{(2)}_{A^{(2)}}=\frac{g_1^4 \rho ^3 \left(4 \rho ^2+21 y^2+19 \rho  y\right)-g_2^2 g_1^2 \rho  (\rho +y)^3 (8 \rho +27 y)+4 g_2^4 (\rho +y)^4 (\rho +4 y)}{y (\rho +y) \left(g_1^4 \rho ^3 (2 \rho +3 y)-g_2^2 g_1^2 \rho  (\rho +y)^2 (4 \rho +3 y)+2 g_2^4
   (\rho +y)^4\right)}.
\end{equation}

\subsection{Solving the third order differential equation for $A^{(2)}$}
\label{GreenF}
In this subsection we solve for the solutions of the homogeneous equation corresponding to (\ref{ThirdOrderEquation}):
\begin{gather}
A^{(2)}_{h1}=a^{(2)}_{h1}(y) C_{8}(t,x), \ A^{(2)}_{h2}=a^{(2)}_{h2}(y) C_{9}(t,x) \text{ and } A^{(2)}_{h3}=a^{(2)}_{h1}(y) C_{10}(t,x),
\end{gather}
where:
\begin{align}
a^{(2)}_{h1}(y)=\frac{y \left(g_1^2 (9-7 y)+4 g_2^2 (y+1) (4 y-5)\right)}{4 \left(g_1^2-4 g_2^2\right) (y+1) \left(g_1^2-g_2^2 (y+1)^2\right)},\\
a^{(2)}_{h2}(y)=\frac{(y-1) y \left(g_1^2-2 g_2^2 (y+1)\right)}{2 \left(g_1^2-4 g_2^2\right) (y+1) \left(g_1^2-g_2^2 (y+1)^2\right)},
\end{align}
\begin{multline}
a^{(2)}_{h3}(y)=\frac{g_2^2 g_1^2 (y+1) \left(y \left(12 y^3-5 y+2\right)-2\right)+g_1^4 \left(2 y \left(6 y^2+3 y-1\right)+1\right)}{6 y^2 (y+1)^2
   \left(g_2 y-g_1+g_2\right) \left(g_2 y+g_1+g_2\right)}\\ \qquad \qquad+\frac{6 g_1^2 y^3 (y+1) \left(g_2^2 \left(2 y^2+y-1\right)+2 g_1^2\right) \log (\frac{y}{y+1})+g_2^4 (y+1)^2}{6 y^2 (y+1)^2
   \left(g_2 y-g_1+g_2\right) \left(g_2 y+g_1+g_2\right)}.
\end{multline}
With this at hand we define the Green function:
\begin{equation}\label{Green}
G(z,y)=u_{h1}(z) a^{(2)}_{h1}(y)+u_{h2}(z) a^{(2)}_{h2}(y)+u_{h3}(z) a^{(2)}_{h3}(y),
\end{equation}
where
\begin{equation}
u_{h3}(z)=\frac{z^4 (z+1)^4}{g_1^2 (3 z+2)-2 g_2^2 (z+1)^2},
\end{equation}
\begin{multline}
u_{h1}(z)=\frac{(z+1)^2 \left(\left(g_1^2+g_2^2\right)g_1^2 \left(12 (z+1)^2 z^4 \log \left(\frac{z}{z+1}\right)\right)\right)}{g_1^2 (9 z+6)-6 g_2^2 (z+1)^2}\\+\frac{(z+1)^2 \left(\left(g_1^2+g_2^2\right)g_1^2 \left((z (2 z+1) (6 z (z+1)-1)+4) z-3\right)\right)}{g_1^2 (9 z+6)-6 g_2^2 (z+1)^2}\\+\frac{(z+1)^2 \left(g_2^2 g_1^2 (z+1) \left(\left(2 z \left(6 z^2+3 z-1\right)-11\right)
   z+9\right)\right)}{g_1^2 (9 z+6)-6 g_2^2 (z+1)^2}\\+\frac{(z+1)^2 \left(g_2^2 g_1^2 (z+1) \left(2 g_2^4 (z+1)^2 (4 z-3)\right)\right)}{g_1^2 (9 z+6)-6 g_2^2 (z+1)^2},
\end{multline}
\begin{multline}
u_{h2}(z)=\frac{(z+1)^2 \left(\left(7 g_1^2+ 12 g_2^2\right)g_1^2 \left(12 (z+1)^2 z^4 \log \left(\frac{z}{z+1}\right)\right)\right)}{6 \left(g_1^2 (3 z+2)-2 g_2^2 (z+1)^2\right)}\\+\frac{(z+1)^2 \left(3 g_2^2 g_1^2 (z+1) \left(\left(8 z \left(6 z^2+3 z-1\right)-27\right)
   z+29\right)\right)}{6 \left(g_1^2 (3 z+2)-2 g_2^2 (z+1)^2\right)}\\+ \frac{(z+1)^2 \left(\left(7 (z (2 z+1) (6 z (z+1)-1)+4) z-27\right)\right)}{6 \left(g_1^2 (3 z+2)-2 g_2^2 (z+1)^2\right)}\\+
   \frac{(z+1)^2 \left(4 g_2^4 (z+1)^2 (16 z-15)\right)}{6 \left(g_1^2 (3 z+2)-2 g_2^2 (z+1)^2\right)}.
\end{multline}
With this at hand we can compute a particular solution
\begin{equation}
A^{(2)}_{p}=-\int dw \ G(y,w) e^{- 2s_p} F(s_p,\tau,w),
\end{equation}
where $e^{-2 s_p} F$ is the RHS inhomogeneity in (\ref{ThirdOrderEquation}).
After integration we get the final expression for $A^{(2)}$. We do not post the result but the computation is straightforward. The remaining background fluctuations, $g^{(2)}$ and $\phi^{(2)}$ are evaluated by use of (\ref{rrIntegrated}) and (\ref{phi2list}) once $A^{(2)}$ is known.
\paragraph{The case of the modulus $\rho$}
In this paragraph we present the results towards the derivation of the Green function of the very last third order differential equation in case only the modulus $\rho$ is turned on. In this case we get the homogeneous solutions of (\ref{ThirdOrderDiffRho}) from:
\begin{multline}
a^{(2)}_{h1}(y)=-\frac{y \left(g_2^2 (\rho +1) \left(-2 \rho  (2 \rho +3)+(3 \rho +5) y^2+\left(3 \rho ^2+\rho -6\right) y\right)+g_1^2 \rho ^2 (4 \rho -(3 \rho +4) y+5)\right)}{(\rho +1)^2 \left(g_1^2 \rho ^2-g_2^2 (\rho +1)^2\right) (\rho +y) \left(g_1^2
   \rho ^2-g_2^2 (\rho +y)^2\right)},
\end{multline}
\begin{equation}
a^{(2)}_{h2}(y)=\frac{(y-1) y \left(g_2^2 (\rho +1) (\rho +y)-g_1^2 \rho ^2\right)}{(\rho +1) \left(g_1^2 \rho ^2-g_2^2 (\rho +1)^2\right) (\rho +y) \left(g_1^2 \rho ^2-g_2^2 (\rho +y)^2\right)},
\end{equation}
\begin{multline}
a^{(2)}_{h3}(y)=-\frac{g_1^4 \rho ^2 \left(12 y^3 (\rho +y) \log \left(\frac{y}{\rho +y}\right)+\rho  \left(\rho ^3+12 y^3+6 \rho  y^2-2 \rho ^2 y\right)\right)}{6 \rho ^4 y^2 (\rho +y)^2 \left(g_2^2 (\rho +y)^2-g_1^2 \rho ^2\right)}\\+\frac{g_2^2 g_1^2 (\rho +y) \left(6 y^3 \left(-\rho ^2+2
   y^2+\rho  y\right) \log \left(\frac{y}{\rho +y}\right)+2 \rho ^4 y\right)}{6 \rho ^4 y^2 (\rho +y)^2 \left(g_2^2 (\rho +y)^2-g_1^2 \rho ^2\right)}\\+\frac{g_2^2 g_1^2 (\rho +y) \left(-2 \rho ^5+12 \rho  y^4-5 \rho ^3 y^2\right)+g_2^4 \rho ^4 (\rho +y)^2}{6 \rho ^4 y^2 (\rho +y)^2 \left(g_2^2 (\rho +y)^2-g_1^2 \rho ^2\right)}.
\end{multline}
To compute the particular solution we obtain :
\begin{equation}
u_{h3}(z)=\frac{z^4 (\rho +z)^4}{2 g_2^2 (\rho +z)^2-g_1^2 \rho  (2 \rho +3 z)},
\end{equation}
\begin{multline}
u_{h1}(z)=\frac{(\rho +1) (\rho +z)^2 \left(g_2^2 g_1^2 (\rho +z) \left(6 \left(\rho ^2-\rho -2\right) z^4 (\rho +z) \log \left(\frac{z}{\rho +z}\right)\right)\right)}{6 \rho ^4 \left(g_1^2 \rho  (2 \rho +3 z)-2 g_2^2 (\rho +z)^2\right)}\\-\frac{(\rho +1) (\rho +z)^2 \left(g_2^2 g_1^2 (\rho +z) \left(\rho  \left(-3 \rho ^4 (2 \rho +1)+6 \left(\rho ^2-\rho -2\right) z^4\right)\right)\right)}{6 \rho ^4 \left(g_1^2 \rho  (2 \rho +3 z)-2 g_2^2 (\rho +z)^2\right)}\\-\frac{(\rho +1) (\rho +z)^2 \left(g_2^2 g_1^2 (\rho +z) \left(\rho  \left(+3 \rho  \left(\rho ^2-\rho
   -2\right) z^3\right)\right)\right)}{6 \rho ^4 \left(g_1^2 \rho  (2 \rho +3 z)-2 g_2^2 (\rho +z)^2\right)}\\-\frac{(\rho +1) (\rho +z)^2 \left(g_2^2 g_1^2 (\rho +z) \left(\rho  \left(\rho ^2 \left(-\rho ^2+\rho +2\right) z^2+\rho ^3 \left(8 \rho ^2+4 \rho -1\right) z\right)\right)\right)}{6 \rho ^4 \left(g_1^2 \rho  (2 \rho +3 z)-2 g_2^2 (\rho +z)^2\right)}\\+\frac{(\rho +1) (\rho +z)^2 \left(g_1^4 \rho ^2 \left(12 z^4 (\rho +z)^2 \log \left(\frac{z}{\rho +z}\right)\right)+g_2^4 \rho ^4 (\rho +1) (4 z-3) (\rho +z)^2\right)}{6 \rho ^4 \left(g_1^2 \rho  (2 \rho +3 z)-2 g_2^2 (\rho +z)^2\right)}\\+\frac{(\rho +1) (\rho +z)^2 \left(g_1^4 \rho ^2 \left(\rho  \left(-3 \rho ^4+12 z^5+18 \rho  z^4+4 \rho
   ^2 z^3-\rho ^3 z^2+4 \rho ^4 z\right)\right)\right)}{6 \rho ^4 \left(g_1^2 \rho  (2 \rho +3 z)-2 g_2^2 (\rho +z)^2\right)},
\end{multline}
\begin{multline}
u_{h2}(z)=-\frac{(\rho +z)^2 \left(-3 g_2^2 g_1^2 (\rho +z) \left(6 \left(\rho ^3-5 \rho -4\right) z^4 (\rho +z) \log \left(\frac{z}{\rho +z}\right)\right)\right)}{}\\-\frac{(\rho +z)^2 \left(-3 g_2^2 g_1^2 (\rho +z) \left(\rho  \left(-\rho ^4 \left(8 \rho ^2+15 \rho +6\right)+6 \left(\rho ^3-5 \rho -4\right) z^4\right)\right)\right)}{}\\-\frac{(\rho +z)^2 \left(-3 g_2^2 g_1^2 (\rho +z) \left(\rho  \left(3 \rho
   \left(\rho ^3-5 \rho -4\right) z^3\right)\right)\right)}{6
   \rho ^4 \left(g_1^2 \rho  (2 \rho +3 z)-2 g_2^2 (\rho +z)^2\right)}\\-\frac{(\rho +z)^2 \left(-3 g_2^2 g_1^2 (\rho +z) \left(\rho  \left(\rho ^2 \left(-\rho ^3+5 \rho +4\right) z^2+\rho ^3 \left(8 \rho ^3+16 \rho ^2+5 \rho -2\right) z\right)\right)\right)}{}\\-\frac{(\rho +z)^2 \left(g_1^4 \rho ^2 \left(12 (3 \rho +4) z^4 (\rho +z)^2 \log \left(\frac{z}{\rho +z}\right)+\rho
    \left(-3 \rho ^4 (4 \rho +5)\right)\right)\right)}{}\\-\frac{(\rho +z)^2 \left(g_1^4 \rho ^2 \left(\rho
    \left(12 (3 \rho +4) z^5+18 \rho  (3 \rho +4) z^4+4 \rho ^2 (3 \rho +4) z^3-\rho ^3 (3 \rho +4) z^2\right)\right)\right)}{}\\-\frac{(\rho +z)^2 \left(g_1^4 \rho ^2 \left(\rho
    \left(4 \rho ^4 (3 \rho +4) z\right)\right)+2 g_2^4 \rho ^4 (\rho +1) (\rho +z)^2 (-6 \rho +2 (3 \rho +5) z-9)\right)}{6
   \rho ^4 \left(g_1^2 \rho  (2 \rho +3 z)-2 g_2^2 (\rho +z)^2\right)},
\end{multline}
that allow us to compute the corresponding Green function from (\ref{Green}). Then we calculate the particular solution by the convolution:
\begin{align}
A^{(2)}_{p}=-\int dw \ G(y,w) F_\rho(w).
\end{align}
The remaining background fluctuations $g^{(2)}$ and $\phi^{(2)}$ are obtained by use of (\ref{rrIntegrated}) and (\ref{phi2list}).
\section{6D  solutions} \label{AppD}

\paragraph{Homogeneous solutions}

In the text we have already given the solutions to the homogeneous differential equations for $\varphi_1$ and $s_1$. For completeness
we give here the solutions to the homogeneous differential equations for the remaining fields:
\begin{eqnarray}
{g_{uv}^{(2)}}_h&=&-a_3\frac{2 r^2 \rho^2 + \rho^4}{2 r^4}+a_1\frac{3 \rho^4}{4 r^4}-a_4(u,v)\bigl(\frac{1}{4}\log(F/r^2)-\frac{ d\rho^4}{32 r^4}-\frac{\rho^2}{4 F}-\frac{( 3r^2+\rho^2) \rho^4}{12 r^4 F}\bigr)+a_7 ,\nonumber\\
{g_{uu}^{(2)}}_h&=&-b_1\frac{1}{2 r^2} + b_2, ~~~~~~{g_{vv}^{(1)}}_h=-c_1\frac{1}{2 r^2} + c_2,\nonumber\\
f^{(2)}_h&=&\frac{\log(F/r^2)}{6 \rho^2 F G}\bigl(72 r^2 F^3 a_1+r^2(F G+ 2  \rho^6)a_4\bigr)+\frac{\log(r/\rho)}{F G}4 (4 + d) r^2 \rho^4 a_3\nonumber\\
   &~&+\rho^2 \frac{3 (4 + d) r^8 - 5 (4 + d) r^6 \rho^2 -
    3 (32 + 11 d) r^4 \rho^4 + (8 - 3 d) r^2 \rho^6 + 2 d \rho^8}{12 F G r^4 } a_3\nonumber\\&~&-\frac{48 r^8 + 72 r^6 \rho^2 + (20 + d) r^4 \rho^4 +
 2 (2 + d) r^2 \rho^6 + d \rho^8}{4  G r^4}a_1\nonumber\\&~&-\frac{48 F^3 G -
 120 F^2 G \rho^2 + (100 + 3 d) F G \rho^4 -
 12 (-4 + d) F^2 \rho^6 - 12 (24 + d) F \rho^8 +
 4 (60 + d) \rho^{10}}{288 F G  r^4}a_4\nonumber\\&~&+\frac{2 r^2 \rho^4}{F G} a_5+\frac{r^2 F^2}{\rho^2 G} a_2 -\frac{r^2}{4\rho^2} a_6,
\end{eqnarray}
where $F = r^2+\rho^2$ and $G= ((4 + d) r^4 +2 (4 + d) r^2 \rho^2 + d \rho^4)$ and $a$, $b$ and $c$ are integration constants that depend only on $u$ and $v$.

\paragraph{Particular solutions}

The particular solution for $s^{(1)}$ is given in the text. The particular solution for the remaining fields is:
\begin{eqnarray}
\varphi^{(2)}_p&=&0,\nonumber\\
{g_{uv}^{(2)}}_p &=&-\log(F/r^2)\frac{8 c \left(-5 \partial_u \rho \partial_v \rho+\rho \partial_u \partial_v \rho  \right)}{\rho ^4}
-c\partial_u \rho \partial_v \rho \frac{(80 r^2+7d \rho^2)F^2-\rho^2(12 r^4-20\rho^4)}{2 r^4 \rho^2 F^2}\nonumber\\&~&+c\partial_u \partial_v \rho \frac{16 r^4+(12+d) r^2 \rho ^2+(4+d) \rho ^4}{2 r^4 \rho  F},\nonumber\\
{g_{uu}^{(2)}}_p &=&-\log(F/r^2)\frac{8 c \left(-3 (\partial_u \rho)^2+\rho \partial_u^2 \rho\right)}{\rho ^4}+\partial_u^2 \rho \frac{4 c(2 r^2+\rho ^2)}{r^2 \rho F}
-(\partial_u \rho)^2\frac{4 c \left(6 r^4+9 r^2 \rho ^2+2 \rho ^4\right)}{r^2 \rho ^2 F^2},\nonumber\\
{g_{vv}^{(2)}}_p &=&-\log(F/r^2)\frac{8 c \left(-3 (\partial_v \rho)^2+\rho \partial_v^2 \rho\right)}{\rho ^4}+\partial_v^2 \rho \frac{4 c(2 r^2+\rho ^2)}{r^2 \rho F}
-(\partial_v \rho)^2\frac{4 c \left(6 r^4+9 r^2 \rho ^2+2 \rho ^4\right)}{r^2 \rho ^2 F^2},\nonumber\\
f^{(2)}_p &=&\log(F/r^2)
\frac{2 c r^2 \left(-9\partial_u \rho \partial_v \rho+\rho  \partial_u \partial_v \rho \right)}{\rho ^6}+\nonumber\\
&c&\partial_u \rho \partial_v \rho \frac{7 G^2 \rho ^6+28 d F \rho ^{12}+144 r^2 \rho ^{12}+G \left(169  F^5-393 F^4 \rho^2 +216 F^3 \rho^{4}+56 F^2  \rho ^6-27 F \rho ^8-29 \rho ^{10}\right)}{6 r^4 \rho ^6 F^2 G}\nonumber\\ &-&c\partial_u \partial_v \rho
\frac{G\left(28 r^8+40 r^6 \rho ^2+6 r^4 \rho ^4+(2+d) r^2 \rho ^6\right)+(4+d) \rho ^8 \left(d F^2-4 r^4+4 r^2 \rho ^2\right)}{6 r^4 \rho ^5 F G}.
\end{eqnarray}

\vskip 1cm
\bibliographystyle{unsrt}
\bibliography{dilaton3.bib}

\begin{thebibliography}{10}

\bibitem{Zamolodchikov}
A.B. Zamolodchikov.
\newblock {Irreversibility of the Flux of the Renormalization Group in a 2D
  Field Theory}.
\newblock {\em JETP Lett.}, 43:730--732, 1986.

\bibitem{Komargodski1}
Zohar Komargodski and Adam Schwimmer.
\newblock {On Renormalization Group Flows in Four Dimensions}.
\newblock {\em JHEP}, 1112:099, 2011.

\bibitem{Komargodski2}
Zohar Komargodski.
\newblock {The Constraints of Conformal Symmetry on RG Flows}.
\newblock {\em JHEP}, 1207:069, 2012.

\bibitem{Schwimmer1}
A.~Schwimmer and S.~Theisen.
\newblock {Spontaneous Breaking of Conformal Invariance and Trace Anomaly
  Matching}.
\newblock {\em Nucl.Phys.}, B847:590--611, 2011.

\bibitem{TheisenMyers}
Henriette Elvang, Daniel~Z. Freedman, Ling-Yan Hung, Michael Kiermaier,
  Robert~C. Myers, et~al.
\newblock {On renormalization group flows and the a-theorem in 6d}.
\newblock {\em JHEP}, 1210:011, 2012.

\bibitem{Hoyos}
Carlos Hoyos, Uri Kol, Jacob Sonnenschein, and Shimon Yankielowicz.
\newblock {The a-theorem and conformal symmetry breaking in holographic RG
  flows}.
\newblock {\em JHEP}, 1303:063, 2013.

\bibitem{Freedman1}
D.Z. Freedman, S.S. Gubser, K.~Pilch, and N.P. Warner.
\newblock {Renormalization group flows from holography supersymmetry and a c
  theorem}.
\newblock {\em Adv.Theor.Math.Phys.}, 3:363--417, 1999.

\bibitem{Freedman2}
O.~DeWolfe, D.Z. Freedman, S.S. Gubser, and A.~Karch.
\newblock {Modeling the fifth-dimension with scalars and gravity}.
\newblock {\em Phys.Rev.}, D62:046008, 2000.

\bibitem{Sinha1}
Robert~C. Myers and Aninda Sinha.
\newblock {Holographic c-theorems in arbitrary dimensions}.
\newblock {\em JHEP}, 1101:125, 2011.

\bibitem{Casini}
H.~Casini and M.~Huerta.
\newblock {A Finite entanglement entropy and the c-theorem}.
\newblock {\em Phys.Lett.}, B600:142--150, 2004.

\bibitem{Ryu}
Shinsei Ryu and Tadashi Takayanagi.
\newblock {Holographic derivation of entanglement entropy from AdS/CFT}.
\newblock {\em Phys.Rev.Lett.}, 96:181602, 2006.

\bibitem{Rattazzi}
Markus~A. Luty, Joseph Polchinski, and Riccardo Rattazzi.
\newblock {The $a$-theorem and the Asymptotics of 4D Quantum Field Theory}.
\newblock {\em JHEP}, 1301:152, 2013.

\bibitem{Bianchi1}
Massimo Bianchi, Daniel~Z. Freedman, and Kostas Skenderis.
\newblock {How to go with an RG flow}.
\newblock {\em JHEP}, 0108:041, 2001.

\bibitem{Bianchi2}
Massimo Bianchi, Oliver DeWolfe, Daniel~Z. Freedman, and Krzysztof Pilch.
\newblock {Anatomy of two holographic renormalization group flows}.
\newblock {\em JHEP}, 0101:021, 2001.

\bibitem{Freedman3}
D.Z. Freedman, S.S. Gubser, K.~Pilch, and N.P. Warner.
\newblock {Continuous distributions of D3-branes and gauged supergravity}.
\newblock {\em JHEP}, 0007:038, 2000.

\bibitem{Girardello1}
L.~Girardello, M.~Petrini, M.~Porrati, and A.~Zaffaroni.
\newblock {Novel local CFT and exact results on perturbations of N=4 superYang
  Mills from AdS dynamics}.
\newblock {\em JHEP}, 9812:022, 1998.

\bibitem{Bajc}
Borut Bajc and Adrian~R. Lugo.
\newblock {On the matching method and the Goldstone theorem in holography}.
\newblock 2013.

\bibitem{Imbimbo}
C.~Imbimbo, A.~Schwimmer, S.~Theisen, and S.~Yankielowicz.
\newblock {Diffeomorphisms and holographic anomalies}.
\newblock {\em Class.Quant.Grav.}, 17:1129--1138, 2000.

\bibitem{Schwimmer2}
A.~Schwimmer and S.~Theisen.
\newblock {Diffeomorphisms, anomalies and the Fefferman-Graham ambiguity}.
\newblock {\em JHEP}, 0008:032, 2000.

\bibitem{Sinha2}
Arpan Bhattacharyya, Ling-Yan Hung, Kallol Sen, and Aninda Sinha.
\newblock {On c-theorems in arbitrary dimensions}.
\newblock {\em Phys.Rev.}, D86:106006, 2012.

\bibitem{Parynha}
Edi Gava, Parinya Karndumri, and K.S. Narain.
\newblock {Two dimensional RG flows and Yang-Mills instantons}.
\newblock {\em JHEP}, 1103:106, 2011.

\bibitem{Shiraz}
Sayantani Bhattacharyya, Veronika~E Hubeny, Shiraz Minwalla, and Mukund
  Rangamani.
\newblock {Nonlinear Fluid Dynamics from Gravity}.
\newblock {\em JHEP}, 0802:045, 2008.

\bibitem{Nishino}
Hitoshi Nishino and Ergin Sezgin.
\newblock {New couplings of six-dimensional supergravity}.
\newblock {\em Nucl.Phys.}, B505:497--516, 1997.

\bibitem{Callan}
Jr. Callan, Curtis~G., Jeffrey~A. Harvey, and Andrew Strominger.
\newblock {Supersymmetric string solitons}.
\newblock 1991.

\bibitem{WittenSI}
Edward Witten.
\newblock {Small instantons in string theory}.
\newblock {\em Nucl.Phys.}, B460:541--559, 1996.

\bibitem{SW}
Nathan Seiberg and Edward Witten.
\newblock {The D1 / D5 system and singular CFT}.
\newblock {\em JHEP}, 9904:017, 1999.

\bibitem{deHaro}
Sebastian de~Haro, Sergey~N. Solodukhin, and Kostas Skenderis.
\newblock {Holographic reconstruction of space-time and renormalization in the
  AdS / CFT correspondence}.
\newblock {\em Commun.Math.Phys.}, 217:595--622, 2001.

\bibitem{kw}
Igor~R. Klebanov and Edward Witten.
\newblock {AdS / CFT correspondence and symmetry breaking}.
\newblock {\em Nucl.Phys.}, B556:89--114, 1999.

\bibitem{hartnoll}
Dionysios Anninos, Sean~A. Hartnoll, and Nabil Iqbal.
\newblock {Holography and the Coleman-Mermin-Wagner theorem}.
\newblock {\em Phys.Rev.}, D82:066008, 2010.

\bibitem{Duff}
M.J. Duff, Hong Lu, and C.N. Pope.
\newblock {Heterotic phase transitions and singularities of the gauge dyonic
  string}.
\newblock {\em Phys.Lett.}, B378:101--106, 1996.

\bibitem{Aharony}
Ofer Aharony and Micha Berkooz.
\newblock {IR dynamics of D = 2, N=(4,4) gauge theories and DLCQ of 'little
  string theories'}.
\newblock {\em JHEP}, 9910:030, 1999.

\bibitem{dps}
Michael~R. Douglas, Joseph Polchinski, and Andrew Strominger.
\newblock {Probing five-dimensional black holes with D-branes}.
\newblock {\em JHEP}, 9712:003, 1997.

\bibitem{witten}
Edward Witten.
\newblock {On the conformal field theory of the Higgs branch}.
\newblock {\em JHEP}, 9707:003, 1997.

\bibitem{Schwimmer3}
A.~Schwimmer and S.~Theisen.
\newblock {Entanglement Entropy, Trace Anomalies and Holography}.
\newblock {\em Nucl.Phys.}, B801:1--24, 2008.

\end{thebibliography}

\end{document}